\documentclass[journal]{IEEEtran}

\usepackage{amsmath}
\usepackage{graphicx}
\usepackage{epstopdf}
\usepackage{epsfig}
\usepackage{amssymb}
\usepackage{float}
\usepackage{booktabs}
\usepackage{lipsum}
\usepackage{multirow}
\usepackage{array}
\usepackage{url}
\usepackage{threeparttable}
\usepackage{lettrine}
\usepackage{enumerate}
\usepackage{makecell}
\usepackage{subfigure}

\allowdisplaybreaks

\usepackage[noend]{algpseudocode}

\usepackage{algorithmicx}
\usepackage{algorithm}

\newtheorem{example}{Example}
\newtheorem{definition}{Definition}
\newtheorem{lemma}{Lemma}
\newtheorem{theorem}{Theorem}

\newtheorem{proposition}{Proposition}



\begin{document}

\title{Active Fault Isolation for Discrete Event Systems}

\author{Lin~Cao,~Shaolong~Shu,~\IEEEmembership{Senior
		Member,~IEEE},  ~and~Feng~Lin,~\IEEEmembership{Fellow,~IEEE}
	\thanks{Manuscript received XXXX XX, 2022; revised XXXX XX, 2022.
		Recommended by XXXX. The work is supported by the National Science Foundation of  China  under  Grants  62073242 and 61773287.}
	\thanks{Lin Cao (e-mail: caoleen@tongji.edu.cn), Shaolong Shu (e-mail: shushaolong@tongji.edu.cn) and Feng~Lin
		(e-mail: flin@wayne.edu) are with the School of Electronics and Information Engineering, Tongji University,
		Shanghai, China. Feng~Lin is also with the Department of
		Electrical and Computer Engineering, Wayne State University,
		Detroit, MI 48202, USA.}}

\maketitle

\begin{abstract}
In practice, we can not only disable some events, but also enforce the occurrence of some events prior to the occurrence of other events by external control. In this paper, we combine these two control mechanisms to synthesize a more powerful supervisor. Here our control goal is to design an isolation supervisor which ensures in the closed-loop system, faults are isolatable in the sense that after a fault occurs, we can determine which type the fault belongs to by observing the output of the closed-loop system. The isolation supervisor starts to work when the occurrence of faults is detected. We then solve the isolation supervisor synthesis problem as follows. For a given discrete event system, we firstly construct a bipartite transition system which includes all feasible isolation supervisors. An isolation supervisor is feasible if it enforces only events that are physically possible. We then develop an algorithm to check whether the synthesis problem is solvable or not. The algorithm can also be used to find a valid isolation supervisor if the synthesis problem is solvable. The method of combining two control mechanisms can be used to synthesize more powerful supervisors for other supervisory control problems of discrete event systems as well.

\end{abstract}


\IEEEpeerreviewmaketitle

\section{Introduction}\label{Sec1}
\lettrine[lines=2]{F}ault diagnosis addresses the problem of identifying and
isolating faults by detecting deviations of the actual behavior of a dynamic system
from its desired behavior, which is an important task in large-scale complex systems \cite{2015ASurvey}. Various approaches have been proposed for
fault diagnosis, including fault trees, expert systems, neural
networks, fuzzy logic, Bayesian networks, and analytical redundancy \cite{1994Real, 2010Asurvey}.
The increasingly stringent requirements
on performance and reliability of complex systems have necessitated the development of sophisticated
and systematic methods for the timely and accurate diagnosis
of faults.

In discrete event systems (DES) \cite{Cassandras2010Introduction}, fault diagnosis has also attracted much attention in the past few decades, which is to determine whether a fault event has occurred and which type it belongs to by observing events generated by a given discrete event system \cite{2013Overview, 2018On}. Diagnosability and isolatability are proposed to solve this problem. Diagnosability says after the occurrence of a fault, we can undoubtedly determine its occurrence with observations \cite{ lin1994diagnosability, SaSe:95, CaBaMo:12, CaMoBaLa:13, 2021Cao, JiHuChKu:01, YoLa:02} and isolatability says we can undoubtedly determine the type of faults with more observations \cite{SaSe:95, JiHuChKu:01, YoLa:02, sampath1996failure, ThTe:05, 2013Diagnosis}. It is obvious that isolatability is stronger than diagnosability.

Since diagnosability is introduced in \cite{lin1994diagnosability, SaSe:95}, it has been investigated extensively in \cite{CaBaMo:12, CaMoBaLa:13, 2021Cao, JiHuChKu:01, YoLa:02}. By constructing a diagnoser, the necessary and sufficient conditions are derived for diagnosability \cite{CaBaMo:12, CaMoBaLa:13, 2021Cao}. The diagnoser can also be used to diagnose the occurrence of faults. In \cite{JiHuChKu:01, YoLa:02}, algorithms with polynomial computational complexity are proposed to verify diagnosability. Diagnosability are extended into distributed discrete event systems and timed discrete event systems in \cite{2005Decentralized, 2013Reliable, 2002Fault, 2010The}.

With the diagnoser, necessary and sufficient conditions are also derived for isolatability \cite{sampath1996failure, ThTe:05, 2013Diagnosis}. However, since isolatability is much stronger than diagnosability, it is difficult to distinguish the type of faults only by observing the occurrence of observable events in practice. Hence, an active isolation approach by integrating control and fault isolation is proposed in \cite{1998Active}. By disabling the occurrence of some events, the behavior of the closed-loop system is reduced and preventing the system from generating some undesired strings with which the ambiguity of fault types can not be clarified. In \cite{2014Active, Stefan2017Optimal, 2016A, 2020Design}, the authors propose an active isolation scheme where a supervisor takes the system away from the uncertain diagnostic states by disabling some controllable events. However, due to the disturbance of uncontrollable events, the above methods of disabling controllable events may not be sufficient to make the system isolatable.

In practice, we can not only disable some events, but also enforce the occurrence of some events prior to the occurrence of some other events by external control \cite{2005Control}. In \cite{2020On}, the authors construct a controller which can enforce and/or disable events to solve the controllability problem of hybrid systems. Combing the two control mechanisms of disablement and enforcement, we can synthesize a more powerful supervisor with which we have more chances to control a given system to be isolatable. In this paper, we focus on how to synthesize such a powerful supervisor.

In order to ensure the normal performance of a given discrete event system, we do not control the system until the occurrence of faults has been detected. Once we determine the occurrence of faults, we will use an isolation supervisor to control the given discrete event system. Our control includes enforcing the occurrence of forcible events and disabling some controllable events. The goal is to ensure the closed-loop system is isolatable. The isolation supervisor synthesis problem is solved as follows. For a given discrete event system, we firstly construct its bipartite transition system $BTS$ which includes all feasible isolation supervisors. We say an isolation supervisor is feasible if it enforces only events that are physically possible. With the $BTS$, we develop an algorithm to remove all states which may block the system from running and calculate all `good' states from which we can control the system to arrive a marked state from which we can distinguish the type of faults. We then derive the necessary and sufficient conditions for the existence of solutions to the isolation supervisor synthesis problem. When the problem is solvable, we propose an algorithm to calculate an isolation supervisor which is valid in the sense that the closed-loop system under its control is isolatable. Furthermore, the valid isolation supervisor ensures the type of faults can be determined fast.

The work in \cite{2014ActiveD} also investigates active diagnosis of DES by enforcing the occurrence of forcible events and an online approach is proposed using $N$-step lookahead windows in \cite{2017N}. However, they model faults as states, not as events and only use the control mechanism of enforcing the occurrence of forcible events.

Comparing with the existing works in \cite{1998Active, 2014Active, Stefan2017Optimal, 2016A, 2020Design,2014ActiveD, 2017N}, our paper is novel in the following aspects. 1) This paper introduces a control framework which does not interfere with the normal performance of a given discrete event system.
2) A more powerful isolation supervisor is synthesized which can not only disable controllable events, but also enforce the occurrence of forcible events. 3) By defining `good' states, algorithms are derived to check whether the supervisor synthesis problem has solutions and calculate an effective solution if possible.

The rest of paper is organized as follows. In Section \ref{Sec2}, we
introduce some necessary notations. In Section \ref{Sec3}, we review diagnosability, isolatability and the diagnoser. In Section \ref{Sec4}, we formally state the isolation supervisor synthesis problem for discrete event systems. In Section \ref{Sec5}, we construct a bipartite transition system which contains all feasible isolation supervisors. In Section \ref{Sec6}, we propose algorithms to check necessary and sufficient conditions for the existence of solutions and obtain a valid isolation supervisor if it exists. In Section \ref{Sec7}, we apply these results to a smart home system. Finally, we conclude the paper in Section \ref{Sec8}.

\section{Background}\label{Sec2}
A discrete event system can be described by an automaton as
\begin{align*}
	G = ( Q, \Sigma, \delta, q_0),
\end{align*}
where $Q$ is the set of discrete states; $\Sigma$ is the set of
discrete events; $q_0$ is the initial state; $\delta:Q \times
\Sigma \rightarrow Q$ is the transition function which describes
the dynamics of the system. We use $(q,\sigma,q')$ to denote a transition such that $\delta(q,\sigma)=q'$. As usual, we extend the transition function to $\delta:Q \times \Sigma^* \rightarrow Q$, where
$\Sigma^*$ denotes the Kleene closure of $\Sigma$. We use $\varepsilon$ to denote the empty string.

We define the active event function $\Gamma_{G}:Q\rightarrow 2^{\Sigma}$ as follows. $\Gamma_{G}(q)=\{\sigma\in \Sigma:\delta(q,\sigma)!\}$ denotes the set of active events from a state $q\in Q$, where $\delta (q, \sigma)!$ means that $\delta (q, \sigma)$ is defined.

For a string $s\in \Sigma^*$, we use $Pr(s)$ to denote the set of all prefixes of $s$ and let $Pr^+(s)=Pr(s)-\{s\}$.

For a string $s\in \Sigma^*$, we use $|s|$ to denote its length. For a set
like $Q$, we use $|Q|$ to denote its cardinality.

The language generated by $G$ is defined as:
\begin{displaymath}
	L(G)=\{s\in \Sigma^{*}:\delta(q_0,s)!\}
\end{displaymath}

The set of observable events is denoted by
$\Sigma_o\subseteq\Sigma$. The set of unobservable events is
denoted by $\Sigma_{uo}=\Sigma-\Sigma_o$. We define the
natural projection $P:\Sigma^*\rightarrow \Sigma_o^*$ as
\begin{align*}
	& P(\varepsilon)=\varepsilon\\
	&P(\sigma)= \left\{
	\begin{array}{ll}
		\sigma & \textrm{if $\sigma\in \Sigma_o$}\\
		\varepsilon & \textrm{if $\sigma\in \Sigma_{uo}$}\\
	\end{array} \right.\\
	& P(s\sigma)=P(s)P(\sigma)\ \mbox{for } s\in \Sigma^*, \sigma\in
	\Sigma
\end{align*}
The inverse projection with respect to $L(G)$ is defined as
$P_{L(G)}^{-1}(t)=\{s\in L(G):P(s)=t\}$.

We assume that not all events in $\Sigma$ are controllable. We denote the set of controllable events as $\Sigma_c \subseteq \Sigma$. The set of uncontrollable events is denoted as $\Sigma_{uc} = \Sigma - \Sigma_c$. We assume that some events are forcible which can be enforced to occur prior to other events. The set of forcible events is denoted as $\Sigma_{en} \subseteq \Sigma$. Note that forcible events can be controllable or uncontrollable.

Let us introduce some more notations. Let $L(G)/s$ denote the
postlanguage of language $L(G)$ after $s$ as
\begin{displaymath}
L(G)/s=\{t\in
\Sigma^*:st\in L(G)\}
\end{displaymath}

Let $L_o(G)$ denote the set of all strings that originate from the initial state $q_0$ and end at an observable event as
\begin{displaymath}
	L_o(G)=\{s  \sigma \in L(G): \sigma \in \Sigma_o\}
\end{displaymath}

The set of all possible states in $G$ reachable from the initial state $q_0$ via strings in $L_o(G)$ which have the same observation $t\in \Sigma_o^*$ is denoted as
\begin{align*}
	SE_G(t)= \{q\in Q: (\exists s\in L_o(G)) t=P(s)\wedge q=\delta(q_0,s)\}
\end{align*}
$SE_G(t)$ is the current state estimate immediately after observing observable string $t$.

\section{Diagnosability, isolatability and diagnoser}\label{Sec3}

In this section, we review results on diagnosability, isolatability and diagnosers of discrete event systems \cite{2013Overview, 2018On, SaSe:95,  sampath1996failure, ThTe:05, 1998Active}.

\subsection{Diagnosability and isolatability}
Let $\Sigma_f\subseteq \Sigma$ denote the set of fault events. Without loss of generality, We assume that $\Sigma_f\subseteq \Sigma_{uo}$ since an observable fault event can be trivially diagnosed. We partition the set of fault events into disjoint sets corresponding to different fault types.
\begin{align*}
	\Sigma_f=\Sigma_{f_1}\cup\cdots \cup\Sigma_{f_k}
\end{align*}
Let $\varPi_f=\{1, 2,\cdots, k\}$ denotes this partition. The occurrence of a fault of type $i$ means that some fault event in $\Sigma_{f_i}$ occurs.

We use $\Psi(\Sigma_{f_i})$ to
denote the set of all strings in $L(G)$ that end with a fault event
$\sigma_f\in \Sigma_{f_i}$ as
\begin{displaymath}
	\Psi(\Sigma_{f_i})=\{s\sigma_{f}\in L(G):\sigma_{f}\in \Sigma_{f_i}\}
\end{displaymath}

We use $\sigma\in s$ to denote $\sigma$ is an event in string $s$.
With a slight abuse of notations, we use $\Sigma_{f_i}\in s$ to denote
the fact that $\sigma_{f}\in s$ for some $\sigma_{f}\in \Sigma_{f_i}$.

As usual, we make the following three assumptions about the system under investigation.
\begin{enumerate}
	\item[A1.]
	Discrete event system $G$ is live, that is, $(\forall q\in Q)
	\Gamma_{G}(q)\not=\emptyset$;
	\item[A2.]
	There exists no cycles of unobservable events in $G$, that is,
	$(\exists n_o\in \mathbb{N})(\forall ust\in L(G)) s\in
	\Sigma^*_{uo}\Rightarrow |s|\leq n_o$, where $\mathbb{N}$ is the set of
	natural numbers.
	\item[A3.]
	Along every string $s$ in $L(G)$, no more than one type of fault events can occur.
\end{enumerate}

Assumptions 1 and 2 are made to ensure that the system always has output (observable events). The third assumption ensures that the isolation problem does not become ambiguous.

Intuitively speaking, a discrete event system $G$ is diagnosable if it
is able to identify the occurrence of fault
events based on observations, within a bounded
number of occurrences of events. It is formally defined as
follows.
\begin{definition}
	A discrete event system $G$ is said to be diagnosable with respect to the projection $P$ and $\Sigma_f$ if the following holds:
	\begin{align*}
		&(\exists n\in \mathbb{N})(\forall s_0\in \Psi(\Sigma_{f}))(\forall s\in L(G)/s_0)(|s|\geq n\Rightarrow D)
	\end{align*}
	where the diagnosability condition $D$ is
	\begin{displaymath}
		(\forall w\in P^{-1}_{L(G)}(P(s_0s)))\Sigma_{f}\in w
	\end{displaymath}
\end{definition}

We say a discrete event system $G$ is isolatable if it
has the ability to identify the specific type of occurred fault
events based on observations, within a bounded
number of occurrences of events. It is formally defined as follows.
\begin{definition}
	A discrete event system $G$ is said to be isolatable with respect to the projection $P$ and the partition $\varPi_f$ on $\Sigma_{f}$ if the following holds:
	\begin{align*}
		&(\forall i\in \varPi_f)(\exists n_i\in \mathbb{N})(\forall s_0\in \Psi(\Sigma_{f_i}))(\forall s\in L(G)/s_0)\\
		&(|s|\geq n_i\Rightarrow D')
	\end{align*}
	where the isolatability condition $D'$ is
	\begin{displaymath}
		(\forall w\in P^{-1}_{L(G)}(P(s_0s)))\Sigma_{f_i}\in w
	\end{displaymath}
\end{definition}

\subsection{Diagnoser}
The diagnoser defined below plays an important role in solving the fault diagnosis problem. It can be thought as an extended observer of $G$ \cite{SaSe:95} and is denoted by
\begin{displaymath}
	G_d=(X,\Sigma_o,\xi,x_{0})
\end{displaymath}
Note that the observer used in this paper is from \cite{SaSe:95}, which is different from the standard observer in \cite{Cassandras2010Introduction}.

The procedure to construct the diagnoser $G_d$ is reviewed as follows.

Step 1: Construct label automaton
\begin{align*}
	G_L=&(\{N,F_1,F_2,\cdots,F_k\},\Sigma_{f},\delta_f,N)
\end{align*}
as shown in Figure 1 where $N$ means no faults have occurred (the system runs normally) and $F_i$ means some faults in $\Sigma_{f_i}$ have occurred. 	
\begin{figure}[htb]
	\centering
	\includegraphics[scale = 0.75]{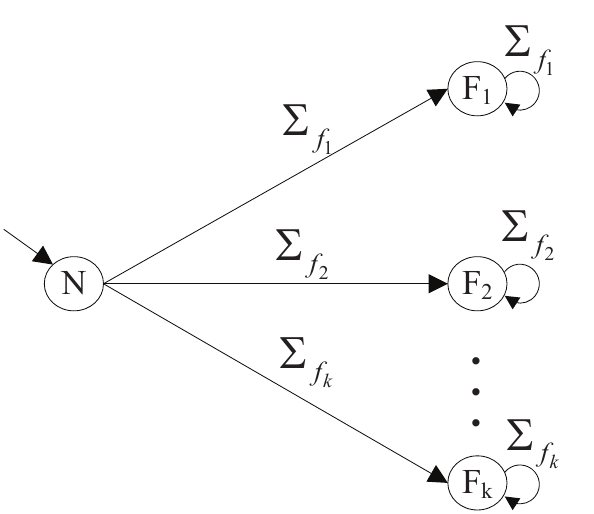}
	\caption{Label automaton $G_L$}
	\label{Fig.1}
\end{figure}

Step 2: Perform parallel composition to obtain a new automaton
\begin{displaymath}
	\hat G=G||G_L=(\hat{Q},\Sigma,\hat{\delta},\hat q_{0}).
\end{displaymath}
$\hat G$ adds labels to each state in $G$ to indicate whether the
fault event $\sigma_{f}$ has occurred or not. Denote
$\hat{q}=(q,l)\in Q \times \{N,F_1,F_2,\cdots,F_k\} = \hat{Q}$. $l=N$ means no faults have occurred. $l=F_i(i\in \{1,2,\cdots,k\})$ means some fault in $\Sigma_{f}$ has
occurred.

Step 3: Construct the observer for $\hat G$ to obtain the diagnoser $G_d$ as
\begin{align*}
	& G_d = (X, \Sigma_o, \xi , x_0) = Ac (2^{\hat{Q}}, \Sigma_o, \xi , (q_0,N)),
\end{align*}
where $Ac(\cdot)$ denotes the accessible part. For an observable event sequence $t\in \Sigma_o^*$, the state $x$ reached in $G_d$ represents the current state estimate as
\begin{displaymath}
	(\forall t\in P(L(G)))\xi(x_0,t)=SE_{\hat G}(t)
\end{displaymath}
The initial state $x_{0}= (q_0,N)$ means no faults occur initially.

For more details, the reader is referred to \cite{SaSe:95}.

\section{Problem statement}\label{Sec4}

Let us continue to consider the diagnoser. With a slight abuse of notations, we use $OSE$ to describe the mapping from the observable strings to the state estimates as
\begin{align*}
OSE:P(L(G))\rightarrow X
\end{align*}
For a given observable string $t\in P(L(G))$, $OSE(t)=SE_{\hat G}(t)=\xi(x_0,t)$.

We divide the state set $X$ into three disjoint subsets: the subset of normal states $X_N$, the set of faulty states $X_{F}$ and the set of uncertain states $X_{U}$ defined as
\begin{align*}
X_N = & \{ x \in X : (\forall (q,l) \in x) l=N\} \\
X_F = & \{ x \in X : (\forall (q,l) \in x) l\in \{F_1,F_2,\cdots,F_k\}\} \\
X_U = & X - (X_N \cup X_F).
\end{align*}
When the diagnoser reaches a state $x\in X_N$, no fault has occurred. When the diagnoser reaches a state $x\in X_F$, a fault has surely occurred. However, when the diagnoser reaches a state $x\in X_{U}$, we can not determinate whether a fault has occurred or not. Let us use $DF:X\rightarrow\{N,F,U\}$ to describe such a mapping as
\begin{align*}
&DF(x)= \left\{
\begin{array}{ll}
N & \textrm{if $x \in X_N$}\\
F & \textrm{if $x \in X_F$}\\
U & \textrm{if $x \in X_U$}\\
\end{array} \right.
\end{align*}

A fault detection agent $A_D$ should tell us whether a fault has occurred or not for every observable string. Hence it is a mapping from the observable string to the set of $\{N,F,U\}$ as
\begin{align*}
A_D:P(L(G))\rightarrow \{N,F,U\}.
\end{align*}
$A_D$ can be calculated as
\begin{align*}
(\forall t\in P(L(G)))A_D(t)=DF\circ OSE(t)=DF(OSE(t))
\end{align*}

We use a fault isolation agent $A_I$ to determine which type of faults has occurred. In order to obtain $A_I$, let us define several subsets as
\begin{align*}
&X_{F_i} = \{ x \in X : (\forall (q,l) \in x) l=F_i\}, i=1,2,\cdots,k
\end{align*}
When the diagnoser reaches a state $x\in X_{F_i}$, it is certain that a fault of the $i$th type has occurred.

Let us use $DI:X\rightarrow\{FU,F_1,F_2,\cdots, F_k\}$ to describe such a mapping as
\begin{align*}
&DI(x)= \left\{
\begin{array}{ll}
F_i & \textrm{if $x \in X_{F_i}$}\\
FU & \textrm{if $x \not\in X_{F_1} \cup ... \cup X_{F_k}$}\\
\end{array} \right.
\end{align*}
The fault isolation agent $A_I$ can be defined as
\begin{align*}
A_I:P(L(G))\rightarrow \{FU,F_1,F_2,\cdots, F_k\}
\end{align*}
and can be calculated as
\begin{align*}
(\forall t\in P(L(G)))A_I(t)=DI\circ OSE(t)=DI(OSE(t))
\end{align*}

\begin{figure}[htb]
	\centering
	\includegraphics[scale = 0.5]{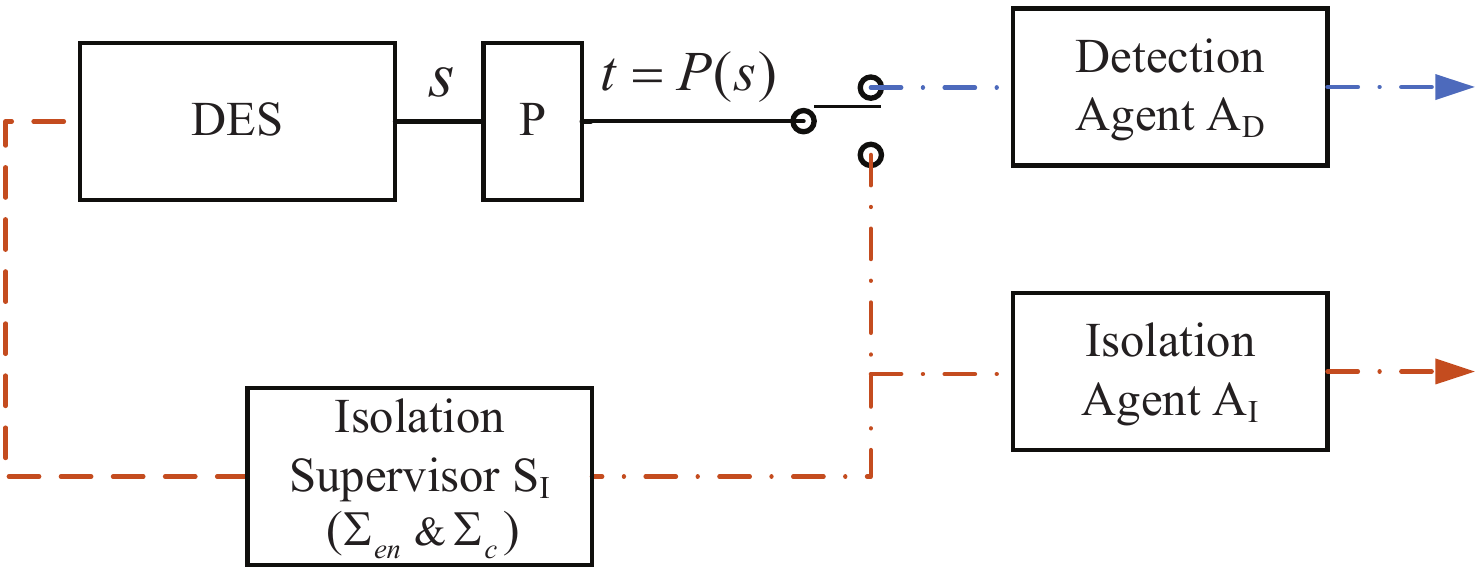}
	\caption{A framework of fault detection and isolation}
	\label{Fig.2}
\end{figure}

We now consider the fault isolation problem for a given discrete event system. Here we control the given system via enforcing the occurrence of some events in $\Sigma_{en}$ and/or disabling the occurrence of some events in $\Sigma_c$. In order to ensure the normal performance of the system, we do not control the system until a fault has occurred and been detected. Therefore, we adopt the fault detection and isolation framework as shown in Figure \ref{Fig.2}.

In Figure 2, the switch is initially connected to the fault detection agent which reports whether a fault has occurred or not so far. Specifically, the fault detection agent outputs the current diagnostic result  $F$, $N$ or $U$ based on the current observation.
Once the fault detection agent outputs $F$, the switch will be turned to the isolation supervisor and the fault isolation agent. The isolation supervisor controls the given system so that the controlled system is isolatable by executing some forcible events and/or disabling some controllable events. The fault isolation agent outputs the type of the occurred fault event as $FU$, $F_1$, $F_2$, $\cdots$, or $F_k$.

Note that $\hat{G}$ generates the same language as  $G$ and its states explicitly tell us whether a fault of type $i$ has occurred or not. Thereafter we will consider $\hat{G}$ instead of $G$.

For a string $s\in L(\hat{G})$, if $\xi(x_0, P(s))\in X_F$, we say it is a diagnosable string since we can determine a fault has occurred via its observation.
We then divide the language $L(\hat{G})$ into two disjoint parts: fault-certain sublanguage, denoted as $L_C(\hat{G})$ and fault-uncertain sublanguage, denoted as $L_{UC}(\hat{G})$. They are defined as
\begin{align*}
&L_C(\hat{G})=\{s\in L(\hat{G}):\xi(x_0, P(s))\in X_F\}\\
&L_{UC}(\hat{G})=L(\hat{G})-L_C(\hat{G})
\end{align*}
Let us further divide the language $L_C(\hat{G})$ into two parts. One is the set of strings ended with an observable event and the other is the set of strings ended with unobservable events. They are defined as
\begin{align*}
&L_{C,o}(\hat{G})=\{s \sigma \in L_C(\hat{G}): \sigma \in \Sigma_o\}\\
&L_{C,uo}(\hat{G})=L_C(\hat{G})-L_{C,o}(\hat{G})
\end{align*}

From Figure \ref{Fig.2}, we know the isolation supervisor will not start until a fault has been determined to occur. We also know that the isolation supervisor issues a new control decision when an observable event is detected. Hence we define the isolation supervisor $S_I$ on the language $P(L_C(\hat{G}))$. For any observation $t\in P(L_C(\hat{G}))$, a control decision should consist of two parts, the event to be enforced to occur and the events to be disabled as, 
\begin{align*}
S_I(t)=<\omega_e(t), \omega_d(t)>
\end{align*}
where $\omega_e(t)\in \Sigma_{en}\cup \{\sim\}$. $\sim$ means no event is enforced to occur when $t$ is observed. $\omega_d(t) \subseteq \Sigma_c$ since uncontrollable events can not be disabled. Note that, while $\omega_e(t)$ is an event, $\omega_d(t)$ is a set of events.

Note that a forcible event can always be enforced to occur prior to the occurrence of other events. When a control decision $<\omega_e(t), \omega_d(t)>$ is issued, the corresponding event $\omega_{e}(t)$ should be executed immediately. It means $\omega_{e}(t)$ should occur after the occurrence of string $s\in L_{C,o}(\hat{G})$ such that $P(s)=t$. \footnote{In traditional supervisory control theory, the occurrence and detection of an observable event are accomplished instantaneously. Hence when an observation $t$ is detected, the occurred string must be one in $L_{C,o}(\hat{G})$ whose observation is $t$.}

When we use an isolation supervisor $S_I$ to control a given system $\hat{G}$, we obtain a closed-loop control system  $S_I/\hat{G}$. 
Note that 
\begin{align*}
L(\hat G)=L_{UC}(\hat G)\cup L_{C,o}(\hat G)\cup L_{C,uo}(\hat G),
\end{align*}
$L(S_I/\hat{G})$ is defined as follows.
\begin{definition}
	The language of closed-loop control system $S_I/\hat{G}$, denoted as $L(S_I/\hat{G})$,  is defined as follows.
	\begin{enumerate}
		\item For all $s\in L_{UC}(\hat{G})$, $s\in L(S_I/\hat{G})$ is always true.
		\item For all $s\in L(S_I/\hat{G})$ and $\sigma\in \Sigma$ such that $s\in L_{C,o}(\hat G)$,
		\begin{align*}
		&s\sigma\in L(S_I/\hat{G})\\
		\Leftrightarrow & s\sigma\in L(\hat{G})\wedge(\sigma=\omega_{e}(P(s))\\
		&\vee(\omega_e(P(s))=\sim\wedge \sigma\notin \omega_d(P(s))))
		\end{align*}
		\item For all $s\in L(S_I/\hat{G})$ and $\sigma\in \Sigma$ such that $s\in L_{C,uo}(\hat G)$, 
		\begin{align*}
		&s\sigma\in L(S_I/\hat{G})\\
		\Leftrightarrow & s\sigma\in L(\hat{G})\wedge\sigma\notin \omega_d(P(s))
		\end{align*}
	\end{enumerate}
\end{definition}

From the definition, we know that all strings in $L_{UC}(\hat{G})$ are in the closed-loop control system $S_I/\hat{G}$ because the isolation supervisor starts to work when a fault is deteced. We also know that forcing can only happen when a string in $L_{C,o}(\hat G)$ occurs because the isolation supervisor issues a control decision when an observable event is observed and forcing should happen immediately. 

We require that an isolation supervisor to be feasible in the sense that it enforces only events that are physically possible in $\hat{G}$. Formally, feasibility is defined as follows.
\begin{definition}
	An isolation supervisor $S_I$ is feasible if
	\begin{align*}
	&(\forall s \in L(S_I/\hat G)\cap L_{C,o}(\hat{G}))\omega_{e}(P(s))\neq \sim\\
	&\Rightarrow s \omega_{e}(P(s))\in L(\hat{G})	
	\end{align*}
\end{definition}

Since diagnosability and isolatability are based on
the assumption that system $\hat{G}$ is live, we want the property of liveness can be held in the controlled system $S_I/\hat{G}$.
\begin{definition}
	Given a discrete event system $\hat{G}$ and an isolation supervisor $S_I$, we say the controlled system $S_I/\hat{G}$ is live if
	\begin{align*}
	(\forall s\in L(S_I/\hat{G})\cap L_C(\hat{G}))(\exists \sigma \in \Sigma)s\sigma\in L(S_I/\hat{G})
	\end{align*}
\end{definition}
Note that $(\forall s \in L_{UC}(\hat{G}))(\exists \sigma \in \Sigma)s\sigma\in L(S_I/\hat{G})$ always holds because $\hat{G}$ is live.

The isolation supervisor is used to control $\hat{G}$ to be isolatable. Hence let us formally define isolatability for the controlled system $S_I/\hat{G}$ as
\begin{definition}
	Given a discrete event system $\hat{G}$ and an isolation supervisor $S_I$, the controlled system $S_I/\hat{G}$ is said to be isolatable if the following statement holds:
	\begin{align*}
	&(\forall i\in \varPi_f)(\exists n_i\in \mathbb{N})(\forall s_0\in \Psi(\Sigma_{f_i}))(\forall s\in L(S_I/\hat{G})/s_0)\\
	&(|s|\geq n_i\Rightarrow D'')
	\end{align*}
	where the condition $D''$ is
	\begin{displaymath}
	(\forall w\in P^{-1}_{L(S_I/\hat{G})}(P(s_0s)))\Sigma_{f_i}\in w
	\end{displaymath}
\end{definition}
Condition $D''$ means that the current state estimate is a $F_i$-state. That is,
\begin{displaymath}
D''\Leftrightarrow SE_{S_I/\hat{G}}(P(s_0s))\in X_{F_i}.
\end{displaymath}

We now consider how to find a feasible isolation supervisor such that the controlled system is isolatable. We formally state it as follows.

\emph{\textbf{Active fault isolation problem for discrete event systems}(AFIP-DES)} Given a discrete event system $\hat{G}$, find a feasible isolation supervisor $S_I$ such that
\begin{enumerate}
	\item $S_I/\hat{G}$ is live;
	\item $S_I/\hat{G}$ is isolatable.
\end{enumerate}

Note that isolatability implies diagnosability. Hence we do not require $S_I/\hat{G}$ to be diagnosable any more.

Let us use an example to to get some intuitive knowledge for the problem.
\begin{example}
	Consider the discrete event system $G$ shown in Figure 3, where
	$\Sigma=\{\sigma_{f_1},\sigma_{f_2},a,o_1,o_2,o_3,o_4\}$,
	$\Sigma_o=\{o_1,o_2,o_3,o_4\}$, $\Sigma_{uo}=\{\sigma_{f_1},\sigma_{f_2},a\}$,
	$\Sigma_{en}=\{o_1,o_2,o_3,a\}$, $\Sigma_{c}=\{o_3\}$
	and $\Sigma_f=\{\sigma_{f_1},\sigma_{f_2}\}$. Clearly, $\Psi(\Sigma_{f_1})=
	\{\sigma_{f_1}\}$, $\Psi(\Sigma_{f_2})=\{\sigma_{f_2}\}$.	
	
	\begin{figure}[htb]
		\centering
		\includegraphics[scale = 0.65]{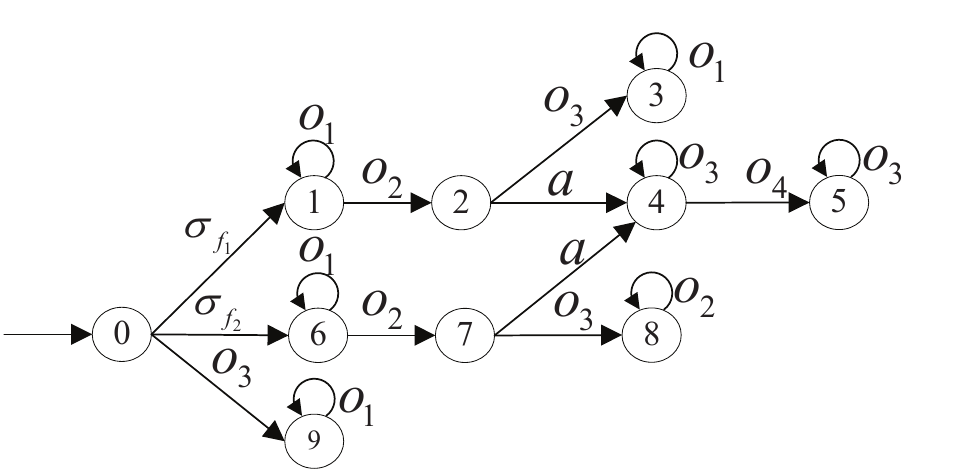}
		\caption{A discrete event system $G$ with $\Sigma_{f_1}=\{\sigma_{f_1}\}$, $\Sigma_{f_2}=\{\sigma_{f_2}\}$, $\Sigma_{o}=\{o_1, o_2, o_3, o_4\}$, $\Sigma_{c}=\{o_3\}$, $\Sigma_{en}=\{o_1, o_2, o_3, a\}$}
		\label{Fig.3}
	\end{figure}

	The corresponding automaton $\hat G$ is shown in Figures 4.
	\begin{figure}[htb]
		\centering
		\includegraphics[scale = 0.65]{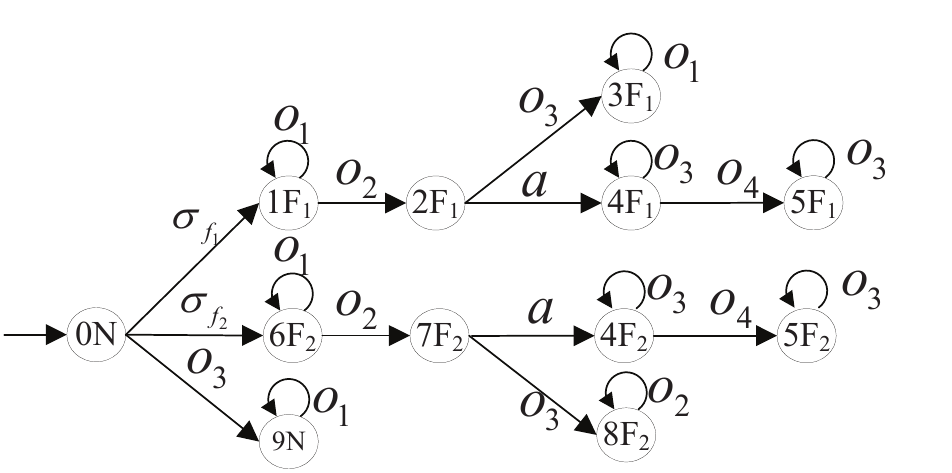}
		\caption{The automaton $\hat{G}$ for $G$}
		\label{Fig.4}
	\end{figure}
	
	Intuitively, $\hat{G}$ is diagnosable because the occurrence of fault events $\sigma_{f_1}$ or $\sigma_{f_2}$ will be diagnosed with the observation of $o_1$ or $o_2$.
	However, $\hat{G}$ is not isolatable. For example, for the observation $o_2o_4{o_3}^*$, there are two corresponding trajectories  $\sigma_{f_1}o_2ao_4{o_3}^*$ and $\sigma_{f_2}o_2ao_4{o_3}^*$. One includes fault event $\sigma_{f_1}$  while the other includes fault event $\sigma_{f_2}$.
	
	Let us try to find an isolation supervisor $S_I$ to ensure that the controlled system $S_I/\hat{G}$ is isolatable. Initially, the isolation supervisor does nothing. Once $o_2$ is observed, the current state estimate is $\{2F_1,7F_2\}$ and the isolation supervisor issues a control policy $S_I(o_2)=<o_3,\emptyset>$ which means event $o_3$ is enforced to occur and no events are disabled. In this case, the system is driven to state $3F_1$ or $8F_2$. Whichever state the system reaches, after one more observation ($o_1$ or $o_2$) is observed, we can determine which type of faults has occurred.
\end{example}

\section{Bipartite transition system}\label{Sec5}
In this section, we investigate how to solve the active fault isolation problem (AFIP-DES). As shown in Figure \ref{Fig.2}, the isolation supervisor does not control the system $\hat G$ until the occurrence of faults is diagnosed. Hence, we have the following proposition.
\begin{proposition}
	For a given system $\hat G$, if there exists one feasible isolation supervisor $S_I$ such that the closed-loop system $S_I/\hat G$ is live and isolatable, then the given system $\hat G$ must be diagnosable.
\end{proposition}
\begin{IEEEproof}	
	Let us prove the result by contradiction. If $G$ is not diagnosable, then there exists a string $s \in L(G)$ of arbitrary length along which a fault has occurred but cannot be diagnosed. Since the fault is not diagnosed, no isolation supervisor $S_I$ will start along this string. Hence, $s \in L(S_I/\hat G)$ for any $S_I$. This implies that $S_I/\hat G$ is not diagnosable, and hence not isolatable. This contradicts the assumption that there exists one feasible isolation supervisor $S_I$ such that the closed-loop system $S_I/\hat G$ is live and isolatable.	
\end{IEEEproof}

Proposition 1 implies that if a given system $\hat G$ is not diagnosable, then the active fault isolation problem (AFIP-DES) has no solutions. From now on, we assume that $\hat G$ is diagnosable.

For a diagnosable system $\hat G$, after a fault occurs, we can determine its occurrence with a finite delay. We define the set of shortest strings along which faults are diagnosed as
\begin{align*}
	STR_{F,0}=&\{s\in L_o(\hat G):\xi(x_0,P(s))\in X_F \\
	&\wedge (\forall s'\in Pr^+(s))\xi(x_0,P(s'))\not\in X_F\}
\end{align*}
Note that any string in $STR_{F,0}$ should end with an observable event.

The set of reachable states in diagnoser via strings in $STR_{F,0}$ is then calculated as
\begin{align*}
X_{F,0}=&\{x:(\exists s\in STR_{F,0})x=\xi(x_0,P(s))\}
\end{align*}

The isolation supervisor $S_I$ starts to work when a string in $STR_{F,0}$ occurs and the diagnoser is in a state belonging to $X_{F,0}$. Our idea is first to construct a bipartite transition system which includes all the feasible isolation supervisors and then remove all the invalid isolation supervisors which can not ensure the closed-loop system to be isolatable. The bipartite transition system should start to run at states in $X_{F,0}$.

Before we construct the bipartite transition system, let us introduce some more notations.

Given a discrete event system $\hat G$, all possible control decisions belong to the Cartesian product of $\Sigma_{en}\cup \{\sim\}$ and $2^{\Sigma_c}$, denoted as
\begin{align*}
\Upsilon=(\Sigma_{en}\cup \{\sim\})\times 2^{\Sigma_c}
\end{align*}

For a state estimate $x$, all the feasible control decisions should be
\begin{align*}
FCD(x)=&\{<\gamma_e,\gamma_d>\in \Upsilon: \gamma_e= \sim\\
&\vee (\forall \hat{q}\in x)\gamma_e\in \Gamma_{\hat{G}}(\hat{q})\}
\end{align*}
Note that when we enforce the occurrence of event $\gamma_e$, $\gamma_e$ is required to be generated from any state $\hat{q}\in x$.

Given a state estimate $x$ and a feasible control decision $<\gamma_e,\gamma_d>$, the system may run to some other states with the occurrence of unobservable events. Those states are called the unobservable reach of $x$. From the unobservable reach, an observable event $\sigma\in \Sigma_o$ continues to occur. Hence the observable reach $OR_{<\gamma_e,\gamma_d>}(x,\sigma)$ from $x$ under the feasible control decision $<\gamma_e,\gamma_d>$ and observation $\sigma$ is calculated as follows.

Case 1: $\gamma_e=\sigma\in \Sigma_o$. Event $\gamma_e=\sigma$ will occur immediately. Hence
\begin{align*}
OR_{<\gamma_e,\gamma_d>}(x,\sigma)=&\{\hat{q}\in \hat{Q}:(\exists \hat{q}'\in x)\hat{q}=\hat{\delta}(\hat{q}',\sigma)\}
\end{align*}

Case 2: $\gamma_e\in \Sigma_{uo}$. Event $\gamma_e$ will occur immediately. Since it is an unobservable event, unobservable events which are not disabled will continue to occur. Hence
\begin{align*}
OR_{<\gamma_e,\gamma_d>}(x,\sigma)=&\{\hat{q}\in \hat{Q}: (\exists \hat{q}''\in x)(\exists s'\in (\Sigma_{uo}-\gamma_d)^*)\\
&\hat{q}'=\hat{\delta}(\hat{q}'',\gamma_e s')\wedge \hat{q}=\hat{\delta}(\hat{q}',\sigma)\}
\end{align*}

Case 3: $\gamma_e= \sim$. In this case, no events are enforced to occur. Some unobservable events which are not disabled will occur. Hence
\begin{align*}
OR_{<\gamma_e,\gamma_d>}(x,\sigma)=&\{\hat{q}\in \hat{Q}: (\exists \hat{q}''\in x)(\exists s'\in (\Sigma_{uo}-\gamma_d)^*)\\
&\hat{q}'=\hat{\delta}(\hat{q}'',s')\wedge \hat{q}=\hat{\delta}(\hat{q}',\sigma)\}
\end{align*}

We can use $OR(\cdot)$ to calculate the state estimate of controlled system $S_I/\hat G$ for any observation $t\in \Sigma_o^*$ as
\begin{proposition}
	For a given system $\hat G$ and an isolation supervisor $S_I$, when an observable event sequence $t=\sigma_1\cdots\sigma_{k-1}\sigma_k$ is observed, the current state estimate can be calculated recursively as
	\begin{align*}
	SE_{S_I/\hat G}(t)=OR_{S_I(\sigma_1\cdots\sigma_{k-1})}(SE_{S_I/\hat G}(\sigma_1\cdots\sigma_{k-1}),\sigma_k)
	\end{align*}
with $SE_{S_I/\hat G}(\varepsilon)=x_0=(q_0,N)$.		
\end{proposition}
\begin{IEEEproof}
	Let $t'=\sigma_1\cdots\sigma_{k-1}$. By the definition of $SE_G(t)$, we have
	\begin{align*}
	SE_{S_I/\hat{G}}(t)=&\{\hat{q}\in \hat{Q}: (\exists s\in L_o(S_I/\hat{G}))\\
	&t=P(s)\wedge \hat{q}=\hat{\delta}(\hat{q}_0,s)\} \tag{5.1}
	\end{align*}
	and
	\begin{align*}
	SE_{S_I/\hat{G}}(t')=&\{\hat{q}\in \hat{Q}: (\exists s\in L_o(S_I/\hat{G}))\\
	&t'=P(s)\wedge \hat{q}=\hat{\delta}(\hat{q}_0,s)\} \tag{5.2}
	\end{align*}
	
	Based on Equations (5.1) and (5.2), we prove the results for the following three cases.
	
	Case 1: $S_I(t')=<\gamma_e,\gamma_d>\wedge\gamma_e=\sigma_k\in \Sigma_o$
	\begin{align*}
	& OR_{<\gamma_e,\gamma_d>}(SE_{S_I/\hat G}(t'),\sigma_k)\\
	=&\{\hat{q}\in \hat{Q}:(\exists \hat{q}'\in SE_{S_I/\hat G}(t'))\hat{q}=\hat{\delta}(\hat{q}',\sigma_k)\} \\
	=&\{\hat{q}\in \hat{Q}:(\exists s\in L_o(S_I/\hat{G}))t'=P(s)\wedge\hat{q}'=\hat{\delta}(\hat{q}_0,s)\\
	&\wedge\hat{q}=\hat{\delta}(\hat{q}',\sigma_k)\} \\
	=&\{\hat{q}\in \hat{Q}: (\exists s\sigma_k\in L_o(S_I/\hat{G}))t=P(s\sigma_k)\wedge\\
	& \hat{q}=\hat{\delta}(\hat{q}_0,s\sigma_k)\}\\
	&\mbox{(By statement 2 in the definition of $S_I/\hat{G}$}\\
	&\mbox{and $\gamma_e=\sigma_k\in \Sigma_o$)}\\
	=&\{\hat{q}\in \hat{Q}: (\exists s'\in L_o(S_I/\hat{G}))t=P(s')\wedge \hat{q}=\hat{\delta}(\hat{q}_0,s')\}\\
	&\mbox{(Let $s'=s\sigma_k$)}\\	
	=&SE_{S_I/\hat{G}}(t)
	\end{align*}
	
	Case 2: $S_I(t')=<\gamma_e,\gamma_d>\wedge\gamma_e\in \Sigma_{uo}$	
	\begin{align*}
	& OR_{<\gamma_e,\gamma_d>}(SE_{S_I/\hat G}(t'),\sigma_k)\\
	=&\{\hat{q}\in \hat{Q}:(\exists \hat{q}''\in SE_{S_I/\hat G}(t'))(\exists s'\in (\Sigma_{uo}-\gamma_d)^*)\\
	&\hat{q}'=\hat{\delta}(\hat{q}'',\gamma_e s')\wedge \hat{q}=\hat{\delta}(\hat{q}',\sigma_k)\} \\
	=&\{\hat{q}\in \hat{Q}:(\exists s\in L_o(S_I/\hat{G}))(\exists s'\in (\Sigma_{uo}-\gamma_d)^*)t'=P(s)\\
	&\wedge\hat{q}''=\hat{\delta}(\hat{q}_0,s)\wedge\hat{q}'=\hat{\delta}(\hat{q}'',\gamma_e s')\wedge\hat{q}=\hat{\delta}(\hat{q}',\sigma_k)\} \\
	=&\{\hat{q}\in \hat{Q}: (\exists s\gamma_e s'\sigma_k\in L_o(S_I/\hat{G}))t=P(s\gamma_es'\sigma_k)\wedge\\
	& \hat{q}=\hat{\delta}(\hat{q}_0,s\gamma_es'\sigma_k)\}\\
	&\mbox{(By statements 2 and 3 in the definition of $S_I/\hat{G}$}\\	
	&\mbox{and $\gamma_e=\sigma_k\in \Sigma_{uo}$)}\\
	=&\{\hat{q}\in \hat{Q}: (\exists s''\in L_o(S_I/\hat{G}))t=P(s'')\wedge \hat{q}=\hat{\delta}(\hat{q}_0,s'')\}\\
	&\mbox{(Let $s''=s\gamma_es'\sigma_k$)}\\	
	=&SE_{S_I/\hat{G}}(t)
	\end{align*}
	
	Case 3: $S_I(t')=<\gamma_e,\gamma_d>\wedge\gamma_e= \sim$
	\begin{align*}
	& OR_{<\gamma_e,\gamma_d>}(SE_{S_I/\hat G}(t'),\sigma_k)\\
	=&\{\hat{q}\in \hat{Q}:(\exists \hat{q}''\in SE_{S_I/\hat G}(t'))(\exists s'\in (\Sigma_{uo}-\gamma_d)^*)\\
	&\hat{q}'=\hat{\delta}(\hat{q}'',s')\wedge \hat{q}=\hat{\delta}(\hat{q}',\sigma_k)\} \\
	=&\{\hat{q}\in \hat{Q}:(\exists s\in L_o(S_I/\hat{G}))(\exists s'\in (\Sigma_{uo}-\gamma_d)^*)\\
	&t'=P(s)\wedge\hat{q}''=\hat{\delta}(\hat{q}_0,s)\wedge\hat{q}'=\hat{\delta}(\hat{q}'',s')\wedge\hat{q}=\hat{\delta}(\hat{q}',\sigma_k)\} \\
	=&\{\hat{q}\in \hat{Q}: (\exists ss'\sigma_k\in L_o(S_I/\hat{G}))t=P(ss'\sigma_k)\\
	& \wedge\hat{q}=\hat{\delta}(\hat{q}_0,ss'\sigma_k)\}\\
	&\mbox{(By Statements 2 and 3 in the definition of $S_I/\hat{G}$)}\\
	=&\{\hat{q}\in \hat{Q}: (\exists s''\in L_o(S_I/\hat{G}))t=P(s'')\wedge \hat{q}=\hat{\delta}(\hat{q}_0,s'')\}\\
	&\mbox{(Let $s''=ss'\sigma_k$)}\\	
	=&SE_{S_I/\hat{G}}(t)
	\end{align*}
\end{IEEEproof}

For an isolation supervisor $S_I$, its feasibility is related with the state estimates as follows.
\begin{proposition}
	For a given system $\hat G$, a given isolation supervisor $S_I$ is feasible if and only if, for any observation $t\in P(L(S_I/\hat G)\cap L_C(\hat{G}))$,
	\begin{align*}
	\omega_e(t)\neq \sim \Rightarrow (\forall \hat q \in SE_{S_I/\hat G}(t))\omega_e(t)\in \Gamma_{\hat G}(\hat q)
	\end{align*}	
\end{proposition}
\begin{IEEEproof}
	An isolation supervisor $S_I$ is feasible, we have
	\begin{align*}
	&(\forall s \in L(S_I/\hat G)\cap L_{C,o}(\hat{G}))\omega_{e}(P(s))\neq \sim\\
	&\Rightarrow s \omega_{e}(P(s))\in L(\hat{G})\\
	\Leftrightarrow &(\forall s\in L(S_I/\hat G)\cap L_{C,o}(\hat{G}))t=P(s)\wedge \omega_{e}(t)\neq \sim\\
	&\Rightarrow s \omega_{e}(t)\in L(\hat{G})\\
	\Leftrightarrow &(\forall s\in L(S_I/\hat G)\cap L_{C,o}(\hat{G}))t=P(s)\wedge \omega_{e}(t)\neq \sim\\
	& \Rightarrow \hat{q}=\hat{\delta}(\hat{q}_0,s)\wedge \hat{\delta}(\hat{q},\omega_{e}(t))!\\
	\Leftrightarrow &(\forall t\in P(L(S_I/\hat G)\cap L_C(\hat{G})))\omega_{e}(t)\neq \sim\Rightarrow\\
	& (\forall \hat q \in SE_{S_I/\hat G}(t))\omega_e(t)\in \Gamma_{\hat G}(\hat q)
	\end{align*}
\end{IEEEproof}

With this knowledge, we construct a bipartite transition system which is inspired by the method proposed in \cite{YiLa16} to solve the standard supervisory control problem for safety and liveness as follows.

The bipartite transition system $BTS$ is a 8-tuple as
\begin{align*}
BTS&=Ac(2^{\hat Q},2^{\hat Q}\times\Upsilon,\delta_{YZ},\delta_{ZY},\Upsilon,\Sigma_o,Y_0,Y_m)\\	&=(Y,Z,\delta_{YZ},\delta_{ZY},\Upsilon,\Sigma_o,Y_0,Y_m)
\end{align*}

$Y\subseteq 2^{\hat Q}$ is the set of $Y$-states. Each $Y$-state is a state estimate, from which the isolation supervisor issues a feasible control decision.
	
$Z\subseteq 2^{\hat Q}\times  \Upsilon$ is the set of $Z$-states. A $Z$-state $z=(z(1),z(2))$ is a doubleton, of which the first element $z(1)$ is the current state estimate and the second element $z(2)$ is the control decision issued at the current state estimate.
	
$\delta_{YZ}: Y\times \Upsilon\rightarrow Z$ is the partial transition function from $Y$-states to $Z$-states, which is defined as follows. For any $y\in Y$, $z\in Z$ and feasible policy $<\gamma_e,\gamma_d>\in FCD(y)$, we have
	\begin{align*}
	&z=\delta_{YZ}(y,<\gamma_e,\gamma_d>)=(y,<\gamma_e,\gamma_d>)
	\end{align*}

$\delta_{ZY}: Z\times\Sigma_o\rightarrow Y$ is the partial transition function from $Z$-states to $Y$-states, which is defined as follows. For any $y\in Y$, $z\in Z$ with $z(2)=<\gamma_e,\gamma_d>$ and $\sigma \in \Sigma_o-\gamma_d$, we have
	\begin{align*}
	&y=\delta_{ZY}(z,\sigma)=OR_{z(2)}(z(1),\sigma)
	\end{align*}

The initial $Y$-states are states in $X_{F,0}$. Hence we define $Y_{0}= \{y_0:y_0\in X_{F,0}\}$.
	
The marked $Y$-state set is defined as $Y_{m}= \{y_m:y_m\in \cup_1^k X_{Fi}\}$. At a marked $Y$-state, the isolation supervisor can determine which type of faults has occurred.
	
We further define $\widetilde{\Sigma}=\Upsilon\cup\Sigma_o$ and $\eta=\delta_{YZ}\cup \delta_{ZY}$.

The bipartite transition system $BTS$ runs as follows. The occurrence of an observable event leads $BTS$ to a state $y \in Y$. $y$ is the state estimate of the observation under current control decision. A new control decision $<\gamma_e,\gamma_d>$ is then laid down. The new control decision drives $BTS$ to a state $z\in Z$ where the current state estimate is saved as $z(1)$  and the control decision is saved as $z(2)$. At state $z$, $BTS$ waits for the occurrence of the next observable event.
	
Let us use an example to show how to construct a bipartite transition system $BTS$ for a given system $\hat G$.
\begin{example}
	Consider again the system $\hat G$ in Figure 4. The automaton $G_d$ is shown in Figure 5.	
	\begin{figure}[htb]
		\centering
		\includegraphics[scale = 0.6]{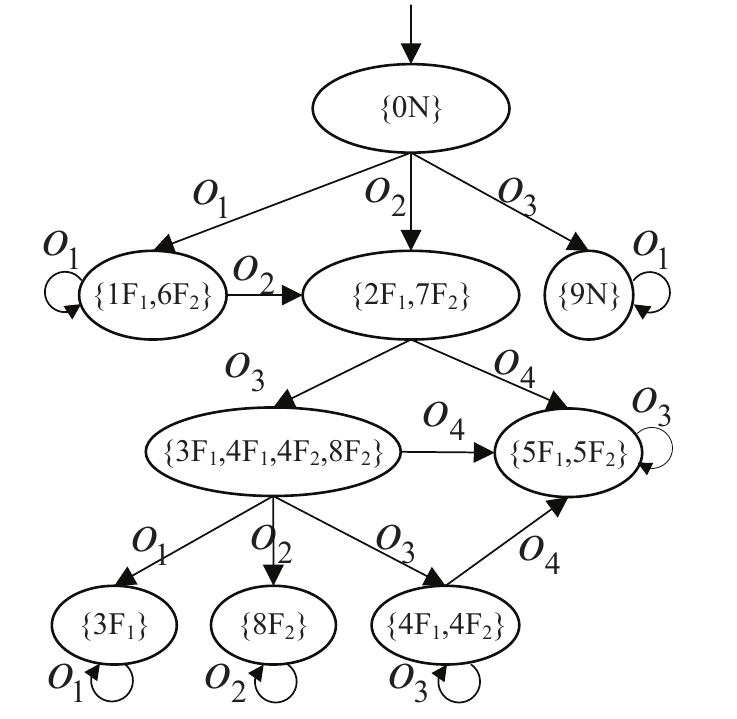}
		\caption{The diagnoser $G_d$ for $\hat G$}
		\label{Fig.5}
	\end{figure}

	From the diagnoser $G_d$, we calculate
	\begin{align*}
	Y_{0}=&\{\{1F_1,6F_2\},\{2F_1,7F_2\}\}\\
	Y_{m}=&\{\{3F_1\},\{8F_2\}\}
	\end{align*}
	
	The bipartite transition system $BTS$ is shown in Figure 6. The initial states in $Y_0$ are marked with blue color and the marked states in $Y_m$ are marked with green color. We use an ellipse to represent a $Y$-state and use a square to represent a $Z$-state.
	
	From initial $Y$-state $\{1F_1,6F_2\}$, there are four feasible control decisions. $<o_1,\emptyset>$ represents enforcing the occurrence of event $o_1$ and disabling nothing. $<o_2,\emptyset>$ represents enforcing the occurrence of event $o_2$ and disabling nothing. $<\sim,\emptyset>$ represents enforcing nothing and disabling nothing. $<\sim,\{o_3\}>$ represents enforcing nothing, but disabling event $o_3$. $BTS$ will reach different $Z$-states $(\{1F_1,6F_2\},<o_1,\emptyset>)$, $(\{1F_1,6F_2\},<o_2,\emptyset>)$, $(\{1F_1,6F_2\},<\sim,\emptyset>)$ and $(\{1F_1,6F_2\},<\sim,\{o_3\}>)$, respectively via these different control decisions.
	
	From $Z$-state $(\{1F_1,6F_2\},<o_1,\emptyset>)$, only one observable event $o_1$ can happen. We have
	\begin{align*}
	&\delta_{ZY}((\{1F_1,6F_2\},<o_1,\emptyset>),o_1)=\{1F_1,6F_2\}
	\end{align*}
	It comes back to the initial $Y$-state $\{1F_1,6F_2\}$. From $(\{1F_1,6F_2\},<o_2,\emptyset>)$, only one observable event $o_2$ can happen, and it leads $BTS$ to $Y$-state $\{2F_1,7F_2\}$.
	
	By the way, we can find all the transitions and obtain the complete bipartite transition system $BTS$ as shown in Figure 6. 
	\begin{figure}[htb]
		\centering
		\includegraphics[scale = 0.65]{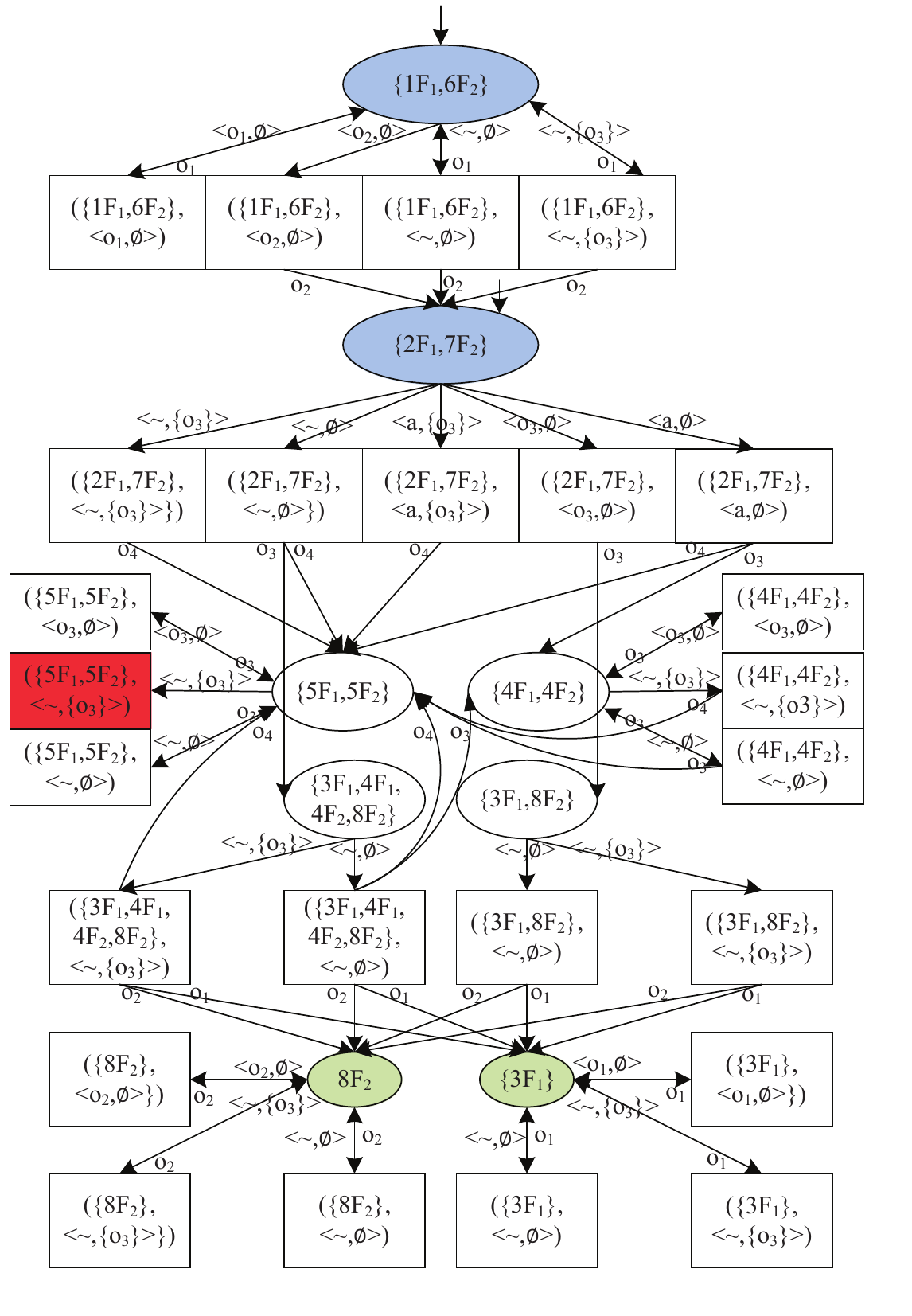}
		\caption{The bipartite transition system $BTS$ for $\hat G$ with $\Sigma_{en}=\{o_1, o_2, o_3, a\}$ and $\Sigma_{c}=\{o_3\}$}
		\label{Fig.6}
	\end{figure}
\end{example}

In $BTS$, every trace consists of observations and control decisions. In order to distinguish observations and control decisions in each $\tilde{s}$ generated by $BTS$, we define two projection functions $M_{\gamma}:(\Upsilon\cup \Sigma_o)^*\rightarrow \Upsilon^*$ and $M_{\Sigma_o}:(\Upsilon\cup \Sigma_o)^*\rightarrow \Sigma_o^*$ inductively as follows:
\begin{enumerate}
	\item $M_{\Upsilon}(\varepsilon)=M_{\Sigma_o}(\varepsilon)=\varepsilon$
	\item For all $\tilde{s}\in (\Upsilon\cup \Sigma_o)^*$ and $\tilde{\sigma}\in (\Upsilon\cup \Sigma_o)$,
	\begin{align*}
	&M_{\Upsilon}(\tilde{s}\tilde{\sigma})= \left\{
	\begin{array}{ll}
	M_{\Upsilon}(\tilde{s})\tilde{\sigma} & \textrm{if $\tilde{\sigma}\in \Upsilon$}\\
	M_{\Upsilon}(\tilde{s}) & \textrm{otherwise}\\
	\end{array} \right.\\
	&M_{\Sigma_o}(\tilde{s}\tilde{\sigma})= \left\{
	\begin{array}{ll}
	M_{\Sigma_o}(\tilde{s})\tilde{\sigma} & \textrm{if $\tilde{\sigma}\in \Sigma_o$}\\
	M_{\Sigma_o}(\tilde{s}) & \textrm{otherwise}\\
	\end{array} \right.
	\end{align*}	
\end{enumerate}

For a given feasible isolation supervisor $S_I$, we consider a string $s\in L(S_I/\hat G)-L_{UD}(\hat G)$. Note that $L_{UD}(\hat G)\subseteq L(S_I/\hat G)$ always holds. We rewrite it as
\begin{align*}
s=s'u_1\sigma_1u_2\sigma_2\cdots u_k\sigma_k u_{k+1}
\end{align*}
where $s'\in STR_{F,0}$, $u_i\in \Sigma_{uo}^*$ and $\sigma_i\in \Sigma_{o}$. Hence $P(s)=P(s')\sigma_1\sigma_2\cdots\sigma_k$.

Starting from $y_0=SE_{S_I/\hat{G}}(P(s'))$, the control decisions issued by $S_I$ for each observation are enumerated as
\begin{align*}
S_I(P(s')), S_I(P(s')\sigma_1), \cdots, S_I(P(s')\sigma_1\cdots\sigma_k)
\end{align*}
Note that we issue a control decision whenever an observable event is observed.

Combining the observations and control decisions, we obtain a sequence of observations and control decisions as
\begin{align*}
\tilde{s}=S_I(P(s'))\sigma_1 S_I(P(s')\sigma_1) , \cdots, \sigma_k S_I(P(s')\sigma_1\cdots\sigma_k )
\end{align*}

For the above $\tilde{s}$, we have
\begin{align*}
M_{\Upsilon}(\tilde{s})=&S_I(P(s'))\sigma_1 S_I(P(s')\sigma_1) , \cdots, \\
&S_I(P(s')\sigma_1\cdots\sigma_k )\\
M_{\Sigma_o}(\tilde{s})=&\sigma_1\sigma_2\cdots\sigma_k
\end{align*}

If $\tilde{s}$ can be generated by $BTS$, we use $Y_{S_I}(y_0,M_{\Sigma_o}(\tilde{s}))$ to denote the last $Y$-state that the sequence of observations and control decisions $\tilde{s}$ visited. It can be calculated recursively as follows.
\begin{align*}
Y_{S_I}(y_0,M_{\Sigma_o}(\tilde{s}))=&Y_{S_I}(y_0,\sigma_1\sigma_2\cdots\sigma_k)\\
=&\delta_{ZY}(\delta_{YZ}(Y_{S_I}(y_0,\sigma_1\sigma_2\cdots\sigma_{k-1}),
\\
&S_I(\sigma_1\sigma_2\cdots\sigma_{k-1})),\sigma_k)
\end{align*}
Initially, we have $Y_{S_I}(y_0,\varepsilon)=y_0$.

Given a $BTS$, we calculate the set of active control decisions generated at $y\in Y$ as
\begin{align*}
C_{BTS}(y)=FCD(y)
\end{align*}

Whenever an observation is observed and $BTS$ reaches a $Y$-state $y$, we select one control decision from $C_{BTS}(y)$. By selecting control decisions for all possible observations, we obtain an isolation supervisor $S_I$. We say that the resulting isolation supervisor $S_I$ is included in $BTS$. For each $Y$-state $y$, $C_{BTS}(y)$ includes all feasible control decisions. Hence all feasible isolation supervisors are included in $BTS$. We also know that all isolation supervisors included in $BTS$ are feasible since all control decisions in $C_{BTS}(y)$ are feasible. For an isolation supervisor $S_I$ included by $BTS$, we have the following theorem.
\begin{theorem}\label{th1}
	Given a system $\hat G$ and an isolation supervisor $S_I$ included in $BTS$. For all
	string $s=s's''\in L(S_I/\hat G)$ such that $s'\in STR_{F,0}$, $Y_{S_I}(y_0,P(s''))$ is the state estimate $SE_{S_I/\hat G}(P(s))$ of the controlled system $S_I/\hat{G}$ after observing $P(s)$, that is,
	\begin{displaymath}
	Y_{S_I}(y_0,P(s''))=SE_{S_I/\hat{G}}(P(s))
	\end{displaymath}
	where $y_0=SE_{S_I/\hat{G}}(P(s'))$.
\end{theorem}
\begin{IEEEproof}
	Let us prove 
	\begin{displaymath}
	Y_{S_I}(y_0,P(s''))=SE_{S_I/\hat{G}}(P(s))
	\end{displaymath}
	by induciton on the length of $P(s'')$. 
		
	Induction Basis: Let $|P(s'')|=0$, that is, $P(s'')=\varepsilon$. We have
	\begin{align*}
	&Y_{S_I}(y_0,\varepsilon)=y_0=SE_{S_I/\hat{G}}(P(s')\varepsilon)
	\end{align*}
	
	Inductive hypothesis: Assume that, for any $P(s'')=\sigma_1\sigma_2\cdots\sigma_k$ such that $|P(s'')|=k\leq n$, we have
	\begin{align*}
	Y_{S_I}(y_0,P(s''))=&Y_{S_I}(y_0,\sigma_1\sigma_2\cdots\sigma_k)\\
	=&SE_{S_I/\hat{G}}(P(s')\sigma_1\sigma_2\cdots\sigma_k)
	\end{align*}
			
	Induction Step: For any $P(s'')=\sigma_1\sigma_2\cdots\sigma_n\sigma_{n+1}$ such that $|P(s'')|=n+1$, we have
	\begin{align*}
	&Y_{S_I}(y_0,\sigma_1\cdots\sigma_{n}\sigma_{n+1})\\
	=&\delta_{ZY}(\delta_{YZ}(Y_{S_I}(y_0,\sigma_0\cdots\sigma_{n}),S_I(\sigma_0\cdots\sigma_{n})),\sigma_{n+1})\\
	=&OR_{S_I(P(s')\sigma_0\cdots\sigma_{n})}(Y_{S_I}(y_0,\sigma_0\cdots\sigma_{n}),\sigma_{n+1})\\
	&\mbox{(By the definition of $\delta_{ZY}(\cdot)$ and $\delta_{YZ}(\cdot)$)}\\
	=&OR_{S_I(P(s')\sigma_0\cdots\sigma_{n})}(SE_{S_I/\hat G}(P(s')\sigma_0\cdots\sigma_{n}),\sigma_{n+1})\\
	&\mbox{(By the inductive hypothesis)}\\
	=&SE_{S_I/\hat{G}}(P(s')\sigma_0\cdots\sigma_{n+1})\\
	&\mbox{(By Proposition 2)}
	\end{align*}
	This completes the proof.
\end{IEEEproof}

\section{Solutions}\label{Sec6}
$BTS$ includes all feasible isolation supervisors. Hence we can solve the active fault isolation problem using $BTS$. We say a feasible isolation supervisor is live if the controlled system $S_I/\hat G$ is live and a feasible isolation supervisor is valid if the controlled system $S_I/\hat G$ is live and isolatable. We use $S(BTS)$, $S_{liv}(BTS)$ and $S_{vld}(BTS)$ to denote the set of all feasible isolation supervisors, all feasible and live isolation supervisors and all feasible and valid isolation supervisors, respectively. We have $S(BTS)\supseteq S_{liv}(BTS) \supseteq  S_{vld}(BTS)$.

We first remove all isolation supervisors in $BTS$ which may cause blocking to obtain a reduced bipartite transition system $BTS_{liv}$.

For a given $BTS$, we say a $Y$-state $y$ is deadlock-free if $C_{BTS}(y)\neq \emptyset$, which means at state $y$, we are able to pick at least one feasible control decision. Otherwise, it is a deadlock.

We consider the deadlock status of a given $Z$-state $z=(y, <\gamma_e, \gamma_d>)$ for the following cases.

Case 1: $\gamma_e\in \Sigma_o$. The $Z$-state $z=(y, <\gamma_e, \gamma_d>)$ is said to be deadlock-free if
\begin{align*}
(\forall \hat q \in y)\delta(\hat q, \gamma_e)!. \tag{6.1}
\end{align*}
Otherwise, it is a deadlock.

Case 2: $\gamma_e\in \Sigma_{uo}$. The $Z$-state $z=(y, <\gamma_e, \gamma_d>)$ is said to be deadlock-free if
\begin{align*}
(\forall \hat q \in y)\delta(\hat q, \gamma_e)! \tag{6.2}
\end{align*}
and
\begin{align*}
&(\forall \hat q'' \in \{\hat q':(\exists \hat q \in y)(\exists s'\in (\Sigma_{uo}-\gamma_d)^*)\hat q'=\delta(\hat q, \gamma_e s')\})\\
&(\exists \sigma\in \Sigma_o-\gamma_d)\delta(\hat q'', \sigma)!. \tag{6.3}
\end{align*}
Otherwise, it is a deadlock.

Case 3: $\gamma_e= \sim$. The $Z$-state $z=(y, <\gamma_e, \gamma_d>)$ is said to be deadlock-free if
\begin{align*}
&(\forall \hat q'' \in \{\hat q':(\exists \hat q \in y)(\exists s'\in (\Sigma_{uo}-\gamma_d)^*)\hat q'=\delta(\hat q, s')\})\\
&(\exists \sigma\in \Sigma_o-\gamma_d)\delta(\hat q'', \sigma)!. \tag{6.4}
\end{align*}
Otherwise, it is a deadlock.

We denote the set of all deadlock $Z$-states as $Z_{DL}$. $Z_{DL}$ can be calculated using Algorithm 1.
\begin{algorithm}
	\caption{calculate all deadlock $Z$-states}
	\label{alg:1}
	\begin{algorithmic}[1]
		\Require
		$BTS$	
		\Ensure
		$Z^{\diamond}$
		\State Set $Z^{\diamond}=\emptyset$
		\For{each $z=(y,<\gamma_e,\gamma_d>)\in Z$}
		\If{$\gamma_e\in \Sigma_o$}
		\State check whether state $z$ is a deadlock by Equation (6.1)
		\If{$z$ is a deadlock}
		\State Update $Z^{\diamond}$ as $Z^{\diamond}\leftarrow Z^{\diamond}\cup\{z\}$
		\EndIf
		\ElsIf{$\gamma_e\in \Sigma_{uo}$}
		\State check whether state $z$ is a deadlock by Equations (6.2) and (6.3)
		\If{$z$ is a deadlock}
		\State Update $Z^{\diamond}$ as $Z^{\diamond}\leftarrow Z^{\diamond}\cup\{z\}$
		\EndIf
		\Else
		\State check whether state $z$ is a deadlock by Equation (6.4)
		\If{$z$ is a deadlock}
		\State Update $Z^{\diamond}$ as $Z^{\diamond}\leftarrow Z^{\diamond}\cup\{z\}$
		\EndIf		
		\EndIf
		\EndFor
		\State Output $Z^{\diamond}$, End	
	\end{algorithmic}
\end{algorithm}

$Z^{\diamond}$ obtained by Algorithm 1 is exactly the set of all deadlock $Z$-states as shown in the following proposition.
\begin{proposition}
	$Z^{\diamond}$ obtained by Algorithm 1 is the set of all deadlock $Z$-states as
	\begin{align*}
	Z^{\diamond}=Z_{DL} 	
	\end{align*}	
\end{proposition}
\begin{IEEEproof}
	It follows from the definition of $Z_{DL}$.
\end{IEEEproof}

In order to construct $BTS_{liv}$, let us introduce the following lemma.
\begin{lemma}
	Given a discrete event system $\hat G$ and a feasible isolation supervisor $S_I$, the controlled system $S_I/\hat G$ is live if and only if all reachable $Z$-states are deadlock-free.
\end{lemma}
\begin{IEEEproof}
	For all $s=s's''\in L(S_I/\hat G)\cap L_C(\hat{G})$ such that $s'\in STR_{F,0}$, we have
	\begin{align*}
	&(\forall s=s's''\in L(S_I/\hat G)\cap L_{C,o}(\hat{G}))s'\in STR_{F,0}\\
	&\wedge y_0=SE_{S_I/\hat{G}}(P(s'))\wedge z=(Y_{S_I}(y_0,P(s'')),S_I(P(s)))\\
	&\wedge \mbox{ $z$ is deadlock-free.}\\
	\Leftrightarrow &(\forall s=s's''\in L(S_I/\hat G)\cap L_{C,o}(\hat{G}))s'\in STR_{F,0}\\
	&\wedge y_0=SE_{S_I/\hat{G}}(P(s'))\wedge ([\omega_e(P(s))\in \Sigma_o\\
	&\wedge(\forall \hat{q}\in Y_{S_I}(y_0,P(s'')))\hat{\delta}(\hat{q},\omega_e(P(s)))!]\\
	&\vee[\omega_e(P(s))\in \Sigma_{uo}\wedge (\forall \hat{q}\in Y_{S_I}(y_0,P(s'')))\\
	&\hat{\delta}(\hat{q},\omega_e(P(s)))!\wedge (\forall \hat q'' \in \{\hat q':(\exists \hat q \in Y_{S_I}(y_0,P(s'')))\\
	&(\exists s_1\in (\Sigma_{uo}-\omega_d(P(s)))^*)\hat q'=\hat \delta(\hat q, \omega_e(P(s)) s_1)\})\\
	&(\exists \sigma\in \Sigma_o-\omega_d(P(s)))\hat \delta(\hat q'', \sigma)!]\vee[\omega_e(P(s))=\sim\\
	&\wedge (\forall \hat q'' \in \{\hat q':(\exists \hat q \in Y_{S_I}(y_0,P(s'')))\\
	&(\exists s_1\in (\Sigma_{uo}-\omega_d(P(s)))^*)\hat q'=\hat \delta(\hat q, s_1)\})\\
	&(\exists \sigma\in \Sigma_o-\omega_d(P(s)))\hat \delta(\hat q'', \sigma)!])\\
	&\mbox{ (Because $z$ is deadlock-free)}\\
	\Leftrightarrow &(\forall s\in L(S_I/\hat G)\cap L_{C,o}(\hat{G}))[\omega_e(P(s))\in \Sigma_o\\
	&\wedge(\forall \hat{q}\in SE_{S_I/\hat{G}}(P(s)))\hat{\delta}(\hat{q},\omega_e(P(s)))!]\\
	&\vee[\omega_e(P(s))\in \Sigma_{uo}\wedge (\forall \hat{q}\in SE_{S_I/\hat{G}}(P(s)))\\
	&\hat{\delta}(\hat{q},\omega_e(P(s)))!\wedge (\forall \hat q'' \in \{\hat q':(\exists \hat q \in SE_{S_I/\hat{G}}(P(s)))\\
	&(\exists s_1\in (\Sigma_{uo}-\omega_d(P(s)))^*)\hat q'=\hat \delta(\hat q, \omega_e(P(s)) s_1)\})\\
	&(\exists \sigma\in \Sigma_o-\omega_d(P(s)))\hat \delta(\hat q'', \sigma)!]\\
	&\vee[\omega_e(P(s))=\sim\wedge (\forall \hat q'' \in \{\hat q':(\exists \hat q \in SE_{S_I/\hat{G}}(P(s)))\\
	&(\exists s_1\in (\Sigma_{uo}-\omega_d(P(s)))^*)\hat q'=\hat \delta(\hat q, s_1)\})\\
	&(\exists \sigma\in \Sigma_o-\omega_d(P(s)))\hat \delta(\hat q'', \sigma)!]\\
	&\mbox{ (By Theorem 1)}\\
	\Leftrightarrow &(\forall s\in L(S_I/\hat G)\cap L_{C,o}(\hat{G}))[\omega_e(P(s))\in \Sigma_o\\
	&\wedge s\omega_e(P(s))\in L(\hat G)]\vee[\omega_e(P(s))\in \Sigma_{uo}\\
	&\wedge s\omega_e(P(s))\in L(\hat G)\wedge (\forall s_1\in (\Sigma_{uo}-\omega_d(P(s)))^*)\\
	&(\exists \sigma\in \Sigma_o-\omega_d(P(s)))s\omega_e(P(s))s_1\sigma\in L(\hat G)]\\
	&\vee[\omega_e(P(s))=\sim\wedge (\forall s_1\in (\Sigma_{uo}-\omega_d(P(s)))^*)\\
	&(\exists \sigma\in \Sigma_o-\omega_d(P(s)))ss_1\sigma\in L(\hat G)]\\
	&\mbox{ (By the definition of $SE_{S_I/\hat{G}}(\cdot)$ and $\hat \delta(\cdot)$)}\\	
	\Leftrightarrow &[(\forall s\in L(S_I/\hat G)\cap L_{C,o}(\hat{G}))\omega_e(P(s))\in \Sigma\\
	&\wedge (\exists \sigma=\omega_e(P(s)))s\sigma\in L(\hat G)]\vee \\
	&[(\forall s\in L(S_I/\hat G)\cap L_{C,o}(\hat{G}))\omega_e(P(s))=\sim\\
	&\wedge(\exists \sigma\in \Sigma-\omega_d(P(s)))s\sigma\in L(\hat G)]\\
	&\vee[(\forall s\in L(S_I/\hat G)\cap L_{C,uo}(\hat{G}))(\exists \sigma\in \Sigma-\omega_d(P(s)))\\
	& s\sigma\in L(\hat G)]\\	
	\Leftrightarrow &[(\forall s\in L(S_I/\hat G)\cap L_{C,o}(\hat{G}))((\omega_e(P(s))\in \Sigma\\
	&\wedge (\exists \sigma=\omega_e(P(s))))\vee (\omega_e(P(s))=\sim \\
	&\wedge(\exists \sigma\in \Sigma-\omega_d(P(s))))) s\sigma\in L(\hat G)]\\
	&\vee [(\forall s\in L(S_I/\hat G)\cap L_{C,uo}(\hat{G}))(\exists \sigma\in \Sigma-\omega_d(P(s)))\\
	& s\sigma\in L(\hat G)]\\
	\Leftrightarrow &[(\forall s\in L(S_I/\hat G)\cap L_{C,o}(\hat{G}))(\exists \sigma\in \Sigma)s\sigma\in L(S_I/\hat G)]\\
	&\vee[(\forall s\in L(S_I/\hat G)\cap L_{C,uo}(\hat{G}))(\exists \sigma\in \Sigma) s\sigma\in L(S_I/\hat G)]\\
	&\mbox{ (By Statements 2, 3 of Definition 3)}\\		
	\Leftrightarrow &(\forall s\in L(S_I/\hat G)\cap L_{C}(\hat{G}))(\exists \sigma\in \Sigma)s\sigma\in L(S_I/\hat G)\\
	\Leftrightarrow &\mbox{$S_I/\hat G$ is live.}
	\end{align*}	
\end{IEEEproof}

For each $Y$-state $y$ in $BTS$, there always exists one feasible control decision which enforces nothing and disables nothing. That is,
\begin{align*}
<\sim, \emptyset>\in C_{BTS}(y)
\end{align*}
always holds. Hence each $Y$-state $y$ in $BTS$ is deadlock-free because $C_{BTS}(y)\neq \emptyset$. With that, we can obtain $BTS _{liv}$ by removing all deadlock $Z$-states in $Z_{DL}$ as
\begin{align*}
BTS _{liv}=&(Y_{liv},Z_{liv},\delta_{YZ,liv},\delta_{ZY,liv},\Upsilon,\Sigma_o,Y_0,Y_m)\\
          =&Ac(Y,Z-Z_{DL},\delta_{YZ},\delta_{ZY},\Upsilon,\Sigma_o,Y_0,Y_m)
\end{align*}

The following theorem shows the correctness of $BTS _{liv}$.
\begin{theorem}
	$BTS _{liv}$ contains all live isolation supervisors, that is
	\begin{align*}	
	S_{liv}(BTS)=S(BTS_{liv})
	\end{align*}
\end{theorem}
\begin{IEEEproof}
	Note that, in $BTS_{liv}$, all $Y$-states and $Z$-states are deadlock-free. Hence each isolation supervisor included in $BTS_{liv}$ is live, that is,
	\begin{align*}	
	(\forall S_I \in S(BTS_{liv})) S_I \ \mbox{is live.} \tag{6.5}
	\end{align*}
	
	Since $BTS_{liv}$ is obtained by removing some control decisions for each $Y$-state in $BTS$, we have
	\begin{align*}	
	S(BTS_{liv})) \subseteq S(BTS)  \tag{6.6}
	\end{align*}
	
	By Equations (6.5) and (6.6), we have
	\begin{align*}	
	S(BTS_{liv})\subseteq S_{liv}(BTS)  \tag{6.7}
	\end{align*}
	
	Now let us show $S_{liv}(BTS)\subseteq S(BTS_{liv})$. From Lemma 1, we know removing deadlock $Z$-states in $Z_{DL}$ does not remove any live isolation supervisor from $BTS$. Hence we have
	\begin{align*}
	S_{liv}(BTS)\subseteq S(BTS_{liv}) \tag{6.8}
	\end{align*}
	
	Combing Equations (6.7) and (6.8), we have
	\begin{align*}	
	S_{liv}(BTS)=S(BTS_{liv})
	\end{align*}
\end{IEEEproof}

\begin{example}
	Let us continue with Example 2. For the bipartite transition system $BTS$ as shown in Figure 6, we use Algorithm 1 to remove all the blocking isolation supervisors. Note that all $Y$-states are deadlock-free. There is only one deadlock $Z$-state $(\{5F_1, 9F_2\},<\sim,\{o_3\}>)$, which is marked with red in Figure 6. 	
\end{example}

For a live isolation supervisor included in $BTS_{liv}$, it has the following property.
\begin{proposition} \label{vldsup}
	Given a discrete event system $\hat G$ and a live isolation supervisor $S_I$ included in $BTS_{liv}$, the controlled system $S_I/\hat G$ is isolatable, that is, $S_I$ is valid, if and only if
	\begin{align*}	
	&(\exists n\in \mathbb{N})(\forall s'\in STR_{F,0})(\forall s=s's''\in L(S_I/\hat G))\\
	&y=SE_{S_I/\hat{G}}(P(s'))\wedge|P(s'')|\geq n\Rightarrow Y_{S_I}(y,P(s''))\in Y_m
	\end{align*}	
\end{proposition}

\begin{IEEEproof}
	By Definition 6, $S_I$ is valid if and only if
	\begin{align*}
		&(\forall i\in \varPi_f)(\exists n'_i\in \mathbb{N})(\forall s'\in \Psi(\Sigma_{f_i}))(\forall s'' \in L(S_I/\hat{G})/s')\\
		&|s''|\geq n'_i\Rightarrow D''\\
		\Leftrightarrow 
		&(\forall i\in \varPi_f)(\exists n'_i\in \mathbb{N})(\forall s'\in \Psi(\Sigma_{f_i}))(\forall s'' \in L(S_I/\hat{G})/s')\\
		&|s''|\geq n'_i\Rightarrow SE_{S_I/\hat{G}}(P(s's''))\in X_{F_i}\\
		&\mbox{(By the condition $D''$)}\\
		\Leftrightarrow 
		&(\forall i\in \varPi_f)(\exists n_i\in \mathbb{N})(\forall s'\in \Psi(\Sigma_{f_i}))(\forall s'' \in L(S_I/\hat{G})/s')\\
		&|P(s'')|\geq n_i\Rightarrow SE_{S_I/\hat{G}}(P(s's''))\in X_{F_i}\\
		&\mbox{(By Assumption 2)}\\
		\Leftrightarrow &(\forall i\in \varPi_f)(\exists n_i\in \mathbb{N})(\forall s'\in STR_{F,0})\\
		&(\forall s=s's''\in L(S_I/\hat{G}))\\
		& |P(s'')|\geq n_i\Rightarrow SE_{S_I/\hat{G}}(P(s))\in X_{F_i}\\
		&\mbox{(By the definition of $STR_{F,0}$ and Proposition 1)}\\
		\Leftrightarrow &(\exists n\in \mathbb{N})(\forall s'\in STR_{F,0})(\forall s=s's''\in L(S_I/\hat{G}))\\
		&y=SE_{S_I/\hat{G}}(P(s'))\wedge|P(s'')|\geq n\\
		& \Rightarrow Y_{S_I}(y,P(s''))\in Y_m\\
		&\mbox{(By the definition of $Y_m$ and Theorem 1)}
	\end{align*}
\end{IEEEproof}

From Proposition \ref{vldsup}, we can see that given a discrete event system $\hat G$  and a live isolation supervisor $S_I$, the controlled system $S_I/\hat G$ is isolatable if and only if the isolation supervisor $S_I$ drives $BTS_{liv}$ into $Y_m$.

For a $BTS_{liv}$, we define the set of observable events defined at $z\in Z_{liv}$ as
\begin{align*}
O_{BTS_{liv}}(z)=\{\sigma\in \Sigma_o: \delta_{ZY,liv}(z,\sigma)!\}
\end{align*}

In $BTS _{liv}$, we are interested in `good' states from which we can find an isolation supervisor driving $BTS _{liv}$ to states in $Y_m$. These  `good' $Y$-states and `good' $Z$-states are defined as follows.
\begin{definition}
	All `good' $Y$-states and $Z$-states are defined recursively as
	
	1. All states in $Y_m$ are `good' states.
	
	2. A $Z$-state $z$ is  `good' if
	\begin{align*}
	&(\forall \sigma\in O_{BTS _{liv}}(z))y=\delta_{ZY,liv}(z,\sigma)\wedge y\ \mbox{is `good'}
	\end{align*}
	
	3. A $Y$-state $y$ is  `good' if
	\begin{align*}
	&(\exists <\gamma_e,\gamma_d>\in \Upsilon)z=\delta_{YZ,liv}(y,<\gamma_e,\gamma_d>)\wedge z\ \mbox{is `good'}
	\end{align*}
\end{definition}

We denote the set of all `good' $Y$-states as $Y_g$ and denote the set of all `good' $Z$-states as $Z_g$. The following algorithm is used to compute all `good' $Y$-states and all `good' $Z$-states and a $Y$-state-based control policy $CP^\diamond$.
\begin{algorithm}
	\caption{Computing all `good' $Y$-states, all `good' $Z$-states and a $Y$-state-based control policy $CP^\diamond$}
	\label{alg:2}
	\begin{algorithmic}[1]
		\Require
		$BTS _{liv}$	
		\Ensure
		$Y^\diamond,Z^\diamond,CP^\diamond$
		\State Set $Y^\diamond=Y_m, Z^\diamond=\emptyset$. For each $y\in Y_m$, arbitrarily select one control decision from $C_{BTS_{liv}}(y)$ as 
		$CP^\diamond(y)$
		\For{each $z\in Z_{liv}-Z^\diamond$}
		\State Check whether state $z$ is `good'
		\If{$z$ is `good'}
		\State Update $Z^\diamond$ as $Z^\diamond\leftarrow Z^\diamond\cup\{z\}$
		\EndIf
		\EndFor
		\For{each $y\in Y_{liv}-Y^\diamond$}
		\State Check whether state $y$ is `good'
		\If{$y$ is `good'}
		\State Update $Y^\diamond$ as
		$Y^\diamond\leftarrow Y^\diamond\cup\{y\}$
		and arbitrarily select one control decision from
		\begin{align*}
		\{ <\gamma_e,\gamma_d>\in \Upsilon:z=\delta_{YZ,liv}(y,<\gamma_e,\gamma_d>)\wedge z\in Z^\diamond\}
		\end{align*}
		as $CP^\diamond(y)$
		\EndIf
		\EndFor	
		\If{$Y^\diamond$ has been updated}
		\State Go to line 2
		\EndIf
		\State Output $Y^\diamond, Z^\diamond, CP^\diamond$, End
	\end{algorithmic}
\end{algorithm}

$Y^\diamond$, $Z^\diamond$ obtained by Algorithm 2 is exactly the set of all `good' $Y$-states and the set of all `good' $Z$-states as shown in the following proposition.
\begin{proposition}
	$Y^\diamond$ and $Z^\diamond$ obtained by Algorithm 2 are  `good' state sets as
	\begin{align*}
	&Y^\diamond=Y_g \wedge Z^\diamond=Z_g	
	\end{align*}	
\end{proposition}
\begin{IEEEproof}
	It follows from the definition of $Y_g$ and $Z_g$.
\end{IEEEproof}

From $Y$-state-based control policy $CP^\diamond$, we can derive an isolation supervisor $S_I^\diamond$ as follows. For any observing string $t\in P(L(S_I^\diamond/\hat{G}))$, we set
\begin{align*}	
&S_I^\diamond(t)= \left\{
\begin{array}{ll}
CP^\diamond(SE_{S_I^\diamond/\hat{G}}(t)) & \textrm{if $SE_{S_I^\diamond/\hat{G}}(t)\in Y_g$}\\
<\sim,\emptyset> & \textrm{otherwise}\\
\end{array} \right.
\end{align*}

With $S_I^\diamond$, let us prove the following proposition.
\begin{proposition}
	For all `good' state $y \in Y_g$, we have
	\begin{align*}
	&y \in Y_g\\
\Leftrightarrow	&(\exists S_I\in BTS_{liv})(\exists n\in \mathbb{N})(\forall s=s's''\in L(S_I/\hat G))\\
&y=SE_{S_I/\hat G}(P(s'))\wedge|P(s'')|\geq n\Rightarrow Y_{S_I}(y,P(s''))\in Y_m
	\end{align*}	
\end{proposition}
\begin{IEEEproof}
We denote the `good' $Y$-state set and the `good' $Z$-state set after the operation of line 1 in Algorithm 2 as $Y_g^0$ and $Z_g^0$, and the `good' $Y$-state set and  the `good' $Z$-state set after the operation of lines 2-9 in Algorithm 2 for the $i$th iteration as $Y_g^i$ and $Z_g^i$. The iteration ends at the $k$th time.

Let us prove the necessity ``$\Rightarrow$''. We firstly prove the following condition always holds by induction on $i$. That is, 
\begin{align*}
&y\in Y_g\\
\Rightarrow &y \in Y_g^{k}\cup Y_g^{k-1}\cup\cdots\cup Y_g^0\\
\Rightarrow &(\exists n\in \mathbb{N})(\forall s=s's''\in L(S_I^\diamond/\hat G))y=SE_{S_I^\diamond/\hat G}(P(s'))\\
&\wedge|P(s'')|\geq n\Rightarrow Y_{S_I^\diamond}(y,P(s''))\in Y_m \tag{6.9}
\end{align*}

Induction Basis: for each $y\in Y_g^0$, we have
\begin{align*}
&y\in Y_m\\
\Rightarrow &Y_{S_I^\diamond}(y,\varepsilon)\in Y_m\\
\Rightarrow &(\exists n=0)(\forall s=s's''\in L(S_I^\diamond/\hat G))y=SE_{S_I^\diamond/\hat G}(P(s'))\wedge\\
&|P(s'')|\geq n\Rightarrow Y_{S_I^\diamond}(y,P(s''))\in Y_m
\end{align*}

Inductive hypothesis: Assume that for $l<k$, we have
\begin{align*}
&y\in Y_g^{l}\cup Y_g^{l-1}\cup\cdots\cup Y_g^0\\
\Rightarrow &(\exists n\in \mathbb{N})(\forall s=s's''\in L(S_I^\diamond/\hat G))y=SE_{S_I^\diamond/\hat G}(P(s'))\wedge\\
&|P(s'')|\geq n\Rightarrow Y_{S_I^\diamond}(y,P(s''))\in Y_m \tag{6.10}
\end{align*}

Induction Step:
For each $y\in Y_g^{l+1}\cup Y_g^{l}\cup Y_g^{l-1}\cup\cdots\cup Y_g^0$, if $y\in Y_g^{l}\cup Y_g^{l-1}\cup\cdots\cup Y_g^0$, Equation (6.10) is satisfied by the inductive hypothesis. Hence we only need to prove when $y\in Y_g^{l+1}$, we have
\begin{align*}
&y\in Y_g^{l+1}\\
\Rightarrow &(\exists <\gamma_e,\gamma_d>\in \Upsilon)z=\delta_{YZ,liv}(y,<\gamma_e,\gamma_d>)\wedge\\
& z\in Z_g^{l+1}\\
&\mbox{(By the definition of `good' $Y$-state)}\\
\Rightarrow &z=\delta_{YZ,liv}(y,CP^\diamond(y))\wedge z\in Z_g^{l+1}\\
&\mbox{(By the definition of $CP^\diamond(\cdot)$)}\\
\Rightarrow &z=\delta_{YZ,liv}(y,CP^\diamond(y))\wedge (\forall \sigma\in O_{BTS _{liv}}(z))\\
&y'=\delta_{ZY,liv}(z,\sigma)\wedge y'\in Y_g^{l}\cup Y_g^{l-1}\cup\cdots\cup Y_g^0\\
&\mbox{(By the definition of `good' $Z$-state)}\\
\Rightarrow &z=\delta_{YZ,liv}(y,CP^\diamond(y))\wedge (\forall \sigma\in O_{BTS _{liv}}(z))\\
&y'=\delta_{ZY,liv}(z,\sigma)\wedge (\exists n\in \mathbb{N})(\forall s=s's''\in L(S_I^\diamond/\hat G))\\
&y'=SE_{S_I^\diamond/\hat G}(P(s'))\wedge |P(s'')|\geq n\\
&\Rightarrow Y_{S_I^\diamond}(y',P(s''))\in Y_m\\
&\mbox{(By Equation (6.10))}\\
\Rightarrow &z=\delta_{YZ,liv}(y,CP^\diamond(y))\wedge (\forall \sigma\in O_{BTS _{liv}}(z))\\
&y'=Y_{S_I^\diamond}(y,\sigma)\wedge(\exists n\in \mathbb{N})(\forall s=s's''\in L(S_I^\diamond/\hat G))\\
&y'=SE_{S_I^\diamond/\hat G}(P(s'))\wedge |P(s'')|\geq n\\
&\Rightarrow Y_{S_I^\diamond}(y',P(s''))\in Y_m\\
&\mbox{(By the definition of $Y_{S_I^\diamond}(\cdot)$)}\\
\Rightarrow &z=\delta_{YZ,liv}(y,CP^\diamond(y))\wedge (\forall \sigma\in O_{BTS _{liv}}(z))\\
&y'=SE_{S_I^\diamond/\hat G}(y,\sigma)\wedge(\exists n\in \mathbb{N})\\
&(\forall s=s's''\in L(S_I^\diamond/\hat G))y'=SE_{S_I^\diamond/\hat G}(P(s'))\wedge\\
& |P(s'')|\geq n\Rightarrow Y_{S_I^\diamond}(y',P(s''))\in Y_m\\
&\mbox{(By Theorem 1)}\\
\Rightarrow &(\exists n\in \mathbb{N})(\forall s=s_1s_2s''\in L(S_I^\diamond/\hat G))\\
&y=SE_{S_I^\diamond/\hat G}(P(s_1))\wedge y'=SE_{S_I^\diamond/\hat G}(y,P(s_2))\wedge\\
& |P(s'')|\geq n\Rightarrow Y_{S_I^\diamond}(y',P(s''))\in Y_m\\
&\mbox{(By the definition of $SE_{S_I^\diamond/\hat G}(\cdot)$)}\\
\Rightarrow &(\exists n\in \mathbb{N})(\forall s=s_1s_2s''\in L(S_I^\diamond/\hat G))\\
&y=SE_{S_I^\diamond/\hat G}(P(s_1))\wedge|P(s_2s'')|\geq n\Rightarrow\\
& Y_{S_I^\diamond}(y,P(s_2s''))\in Y_m\\
\Rightarrow &(\exists n\in \mathbb{N})(\forall s=s's''\in L(S_I^\diamond/\hat G))y=SE_{S_I^\diamond/\hat G}(P(s'))\\
&\wedge|P(s'')|\geq n\Rightarrow Y_{S_I^\diamond}(y,P(s''))\in Y_m\\
&\mbox{(Let $s'=s_1,s''=s_2s''$)}
\end{align*}
	
With that, Equation (6.9) is proved successfuly. Since $S_I^\diamond\in BTS_{liv}$, we have
\begin{align*}
&y\in Y_g\\
\Rightarrow &(\exists S_I\in BTS_{liv})(\exists n\in \mathbb{N})(\forall s=s's''\in L(S_I/\hat G))y=\\
&SE_{S_I/\hat G}(P(s'))\wedge|P(s'')|\geq n\Rightarrow Y_{S_I}(y,P(s''))\in Y_m
\end{align*}

Let us next prove the sufficiency ``$\Leftarrow$'' by contradiction. Suppose ``$\Leftarrow$'' is not true, that is, there exists a $Y$-state $y$ such as
\begin{align*}
&y\not \in Y_g\\
&\wedge ((\exists S_I\in BTS_{liv})(\exists n\in \mathbb{N})(\forall s=s's''\in L(S_I/\hat G))\\
&y=SE_{S_I/\hat G}(P(s'))\wedge|P(s'')|\geq n\Rightarrow Y_{S_I}(y,P(s''))\in Y_m)
\end{align*}

We know that from any state $Y$-state $y$ in $BTS_{liv}$, there exists an isolation supervisor $S_I$ which drives $BTS_{liv}$ into $Y_m$ within finite observations. Let $n=0,\cdots,k$ to denote the number of observations for each sequence of observations and control decisions in $BTS_{liv}$ under the control of supervisor $S_I$. The maximal number of observations is $k$. The set of $Y$-states visited by $BTS_{liv}$  from $y$ to states in $Y_m$ is denoted as $Y_c$. Based on the number of observations with which $BTS_{liv}$ reaches $Y_m$, we divide $Y_c$ into a serial of subsets as $Y_c^1, \cdots, Y_c^i, \cdots, Y_c^{k-1}, Y_c^k$. As shown in the following figure, the supervisor $S_I$ drives $BTS_{liv}$ from $Y_c^{i+1}$ to $Y_c^{i}$, $Y_c^{i-1}$, $\cdots$, $Y_c^{1}$.
\begin{figure}[htb]
	\centering
	\includegraphics[scale = 0.7]{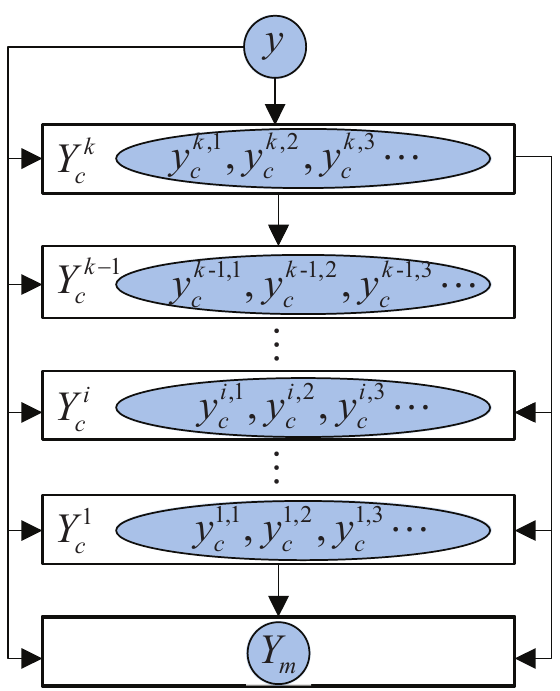}
	\caption{The trajectory from $y$ to $Y_m$ under the control of supervisor $S_I$}
	\label{Fig.7}
\end{figure}

By the definition of `good' $Y$-states and $Z$-states, we know that states in $Y_c^1$ are `good' and then know states in $Y_c^2$, $\cdots$, $Y_c^k$ are all `good' states. Finally, we know $y$ is a `good' $Y$-state, which contradicts $y \not \in Y_g$.
\end{IEEEproof}

With the set of `good' states, we have the following theorem to check the existence of solutions of AFIP-DES and find a valid solution as follows.
\begin{theorem}
	AFIP-DES is solvable if and only if $Y_0\subseteq Y_g$.
	When AFIP-DES is solvable, $S_I^\diamond$ is a solution as
	\begin{align*}	
	S_I^\diamond\in S_{vld}(BTS).
	\end{align*}
\end{theorem}
\begin{IEEEproof}

	\begin{align*}	
		&Y_0\subseteq Y_g\\
		\Leftrightarrow	&(\forall y\in Y_0)y\in Y_g\\
		\Leftrightarrow	&(\forall y\in X_{F,0})y\in Y_g\\
		&\mbox{(By the definition of $Y_0$)}\\
		\Leftrightarrow	&(\forall s'\in STR_{F,0})y=\xi(x_0,P(s'))\wedge y\in Y_g\\
		&\mbox{(By the definition of $X_{F,0}$)}\\
		\Leftrightarrow	&(\forall s'\in STR_{F,0})y=SE_{\hat G}(P(s'))\wedge y\in Y_g\\
		&\mbox{(By the property of $G_d$)}\\
		\Leftrightarrow	&(\forall s'\in STR_{F,0})y=SE_{\hat G}(P(s'))\wedge (\exists S_I^y\in BTS_{liv})\\
		&(\exists n\in \mathbb{N})(\forall w=w'w''\in L(S_I^y/\hat G))\\
		&y=SE_{S_I^y/\hat G}(P(w'))\wedge|P(w'')|\geq n \\
		& \Rightarrow Y_{S_I^y}(y,P(w''))\in Y_m\\
		&\mbox{(By Proposition 7)}\\
		\Leftrightarrow	&(\exists S_I\in BTS_{liv})(\forall s'\in STR_{F,0})y=SE_{S_I/\hat G}(P(s'))\\
		&\wedge (\exists S_I^y\in BTS_{liv})(\exists n\in \mathbb{N})(\forall w=w'w''\in L(S_I^y/\hat G))\\
		&y=SE_{S_I^y/\hat G}(P(w'))\wedge |P(w'')|\geq n\Rightarrow\\
		& Y_{S_I^y}(y,P(w''))\in Y_m\\	
		&\mbox{(Because $SE_{\hat G}(P(s'))=SE_{S_I/\hat G}(P(s'))$ and}\\
		&\mbox{let $S_I(s'w'')=S_I^y(w'w'')$)}\\
		\Leftrightarrow	&(\exists S_I\in BTS_{liv})(\exists n\in \mathbb{N})(\forall s'\in STR_{F,0})\\
		&(\forall s=s'w''\in L(S_I/\hat G))y=SE_{S_I/\hat G}(P(s'))\\
		&\wedge |P(w'')|\geq n\Rightarrow Y_{S_I}(y,P(w''))\in Y_m\\
		&\mbox{(Because $L(S_I/\hat G)/s'=L(S_I^y/\hat G)/w'=L(S_I^y/\hat G,y)$)}\\	
		\Leftrightarrow	&(\exists S_I\in BTS_{liv})(\exists n\in \mathbb{N})(\forall s'\in STR_{F,0})\\
		&(\forall s=s's''\in L(S_I/\hat G))y=SE_{S_I/\hat G}(P(s'))\\
		& \wedge |P(s'')|\geq n\Rightarrow Y_{S_I}(y,P(s''))\in Y_m\\
		&\mbox{(Let $s''=w''$)}\\	
		\Leftrightarrow	&(\exists S_I\in BTS_{liv}) S_I\ \mbox{is valid}\\
		&\mbox{(By Proposition 5)}\\
		\Leftrightarrow &\mbox{AFIP-DES is solvable.}	
	\end{align*}
	
	Suppose that AFIP-DES is solvable, let us now prove $S_I^\diamond$ is a solution. Since $S_I^\diamond$ is live, we only need to prove $S_I^\diamond$ is valid as follows. 	
	\begin{align*}	
	&Y_0\subseteq Y_g\\
	\Rightarrow	&(\forall y\in Y_0)y\in Y_g\\
	\Rightarrow	&(\forall s'\in STR_{F,0})y=SE_{\hat G}(P(s'))\wedge y\in Y_g\\
	&\mbox{(By the property of $Y_0$, $X_{F,0}$ and $G_d$)}\\
	\Rightarrow	&(\forall s'\in STR_{F,0})y=SE_{\hat G}(P(s'))\wedge (\exists n\in \mathbb{N})\\
	&(\forall w=w'w''\in L(S_I^\diamond/\hat G))y=SE_{S_I^\diamond/\hat G}(P(w'))\\
	&\wedge|P(w'')|\geq n\Rightarrow Y_{S_I^\diamond}(y,P(w''))\in Y_m\\
	&\mbox{(By Equation (6.9))}\\
	\Rightarrow	&(\forall s'\in STR_{F,0})y=SE_{S_I^\diamond/\hat G}(P(s'))\wedge (\exists n\in \mathbb{N})\\
	&(\forall w=w'w''\in L(S_I^\diamond/\hat G))y=SE_{S_I^\diamond/\hat G}(P(w'))\\
	&\wedge |P(w'')|\geq n\Rightarrow Y_{S_I^\diamond}(y,P(w''))\in Y_m\\
	&\mbox{(Because $SE_{\hat G}(P(s'))=SE_{S_I^\diamond/\hat G}(P(s'))$)}\\
	\Rightarrow	&(\exists n\in \mathbb{N})(\forall s'\in STR_{F,0})\\
	&(\forall s=s'w''\in L(S_I^\diamond/\hat G))y=SE_{S_I^\diamond/\hat G}(P(s'))\\
	&\wedge |P(w'')|\geq n\Rightarrow Y_{S_I^\diamond}(y,P(w''))\in Y_m\\
	&\mbox{(Because $L(S_I^\diamond/\hat G)/s'=L(S_I^\diamond/\hat G)/w'=L(S_I^\diamond/\hat G,y)$)}\\	
	\Rightarrow	&(\exists n\in \mathbb{N})(\forall s'\in STR_{F,0})\\
	&(\forall s=s's''\in L(S_I^\diamond/\hat G))y=SE_{S_I^\diamond/\hat G}(P(s'))\\
	& \wedge |P(s'')|\geq n\Rightarrow Y_{S_I^\diamond}(y,P(s''))\in Y_m\\
	&\mbox{(Let $s''=w''$)}\\	
	\Rightarrow	&S_I^\diamond\ \mbox{is valid}\\
	&\mbox{(By Proposition 5)}
	\end{align*}
\end{IEEEproof}

Let us use an example to illustrate these results.
\begin{example}
	With $BTS _{liv}$ obtained by Example 3, we use Algorithm 2 to calculate $Y_g$, $Z_g$ and $CP^\diamond$. 	
	
	Initially, we set $Y_g=Y_m=\{\{3F_1\},\{8F_2\}\}$, $Z_g=\emptyset$,  and arbitrarily choose one control decision for each $y\in Y_m$ as $CP^\diamond(y)\in C_{BTS_{liv}}(y)$. We have
	\begin{align*}
	CP^\diamond(\{8F_2\})=&<\sim,\emptyset>,\\
	CP^\diamond(\{3F_1\})=&<\sim,\emptyset>.
	\end{align*}
	
	We then updated $Z_g$ and $Y_g$ as
	\begin{align*}
	Z_g=&\{(\{8F_2\},<\sim,\{o_3\}>), (\{8F_2\},<\sim,\emptyset>),\\
	&(\{8F_2\},<o_2,\emptyset>),(\{3F_1\},<\sim,\{o_3\}>), \\
	&(\{3F_1\},<\sim,\emptyset>), (\{3F_1\},<o_1,\emptyset>),\\
	&(\{3F_1,8F_2\},<\sim,\{o_3\}>), (\{3F_1,8F_2\},<\sim,\emptyset>)\},\\
	Y_g=&\{\{3F_1,8F_2\},\{3F_1\},\{8F_2\}\}.
	\end{align*}
	
	For the added `good' $Y$-state $\{3F_1,8F_2\}$, we determine its control decision as
	\begin{align*}
	CP^\diamond(\{3F_1,8F_2\})=<\sim,\emptyset>.
	\end{align*}
	
	Finally, we obtain all `good' $Y$-states, all `good' $Z$-states as
	\begin{align*}
	Z_g=&\{(\{1F_1,6F_2\},<o_1,\emptyset>),(\{1F_1,6F_2\},<o_2,\emptyset>), \\
	&(\{1F_1,6F_2\},<\sim,\emptyset>), (\{1F_1,6F_2\},<\sim,\{o_3\}>),\\
	&(\{2F_1,7F_2\},<o_3,\emptyset>), (\{3F_1,8F_2\},<\sim,\emptyset>),\\
	&(\{3F_1,8F_2\},<\sim,\{o_3\}>),(\{8F_2\},<\sim,\{o_3\}>),\\
	&(\{8F_2\},<\sim,\emptyset>), (\{8F_2\},<o_2,\emptyset>),\\
	&(\{3F_1\},<\sim,\{o_3\}>), (\{3F_1\},<\sim,\emptyset>),\\
	&(\{3F_1\},<o_1,\emptyset>)\},\\
	Y_g=&\{\{1F_1,6F_2\},\{2F_1,7F_2\},\{3F_1,8F_2\},\{3F_1\},\{8F_2\}\}.
	\end{align*}
	which are marked with blue in Figure 8. The control policy $CP^\diamond$ is also marked with red in Figure 8. For example, we choose control decision $<o_2,\emptyset>$ for $Y$-state $\{1F_1,6F_2\}$.
	\begin{figure}[htb]
		\centering
		\includegraphics[scale = 0.65]{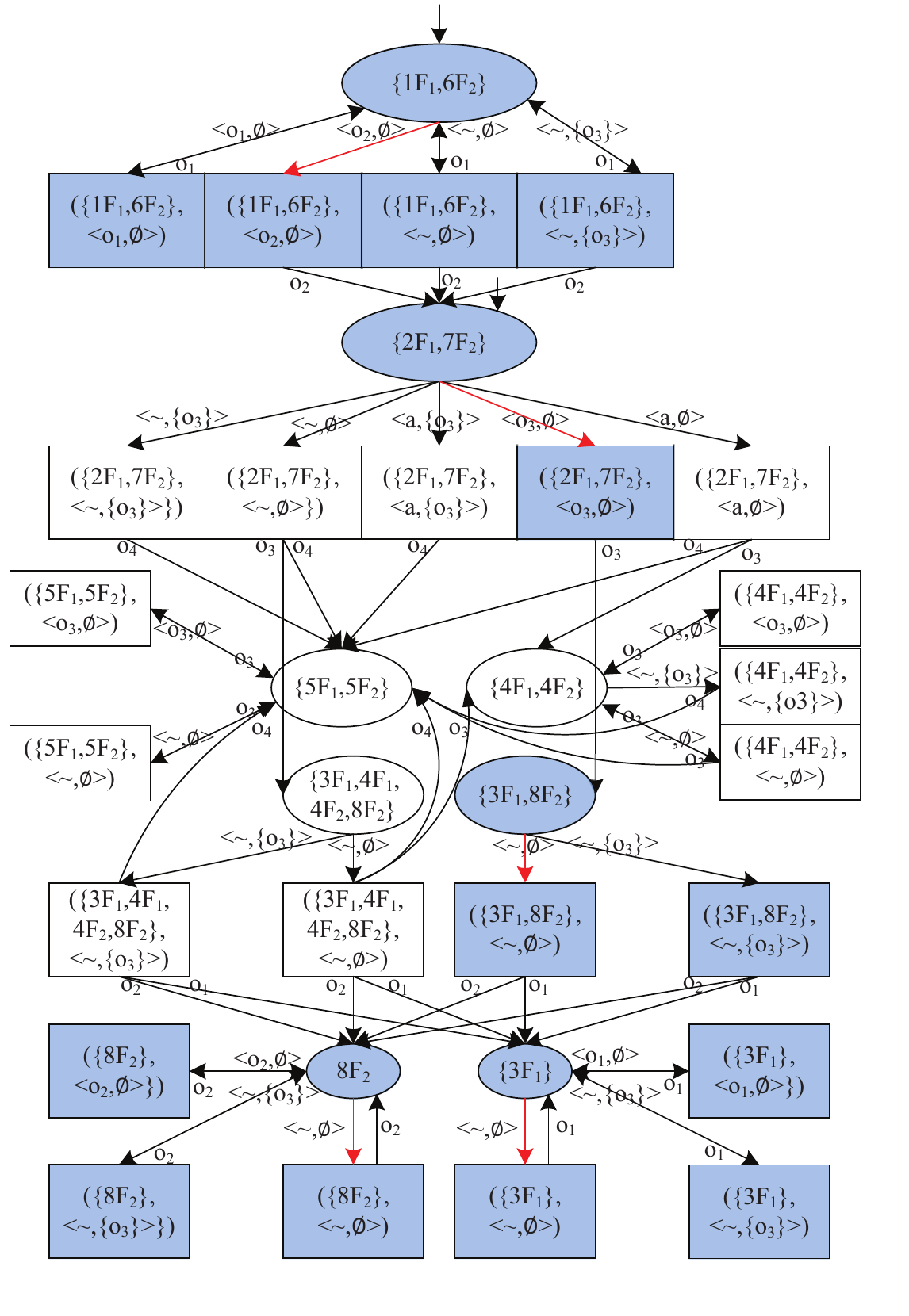}
		\caption{Computing all `good' $Y$-states, all `good' $Z$-states and a $Y$-state-based control policy $CP^\diamond$ for system in Figure 6.}
		\label{Fig.8}
	\end{figure}
	
	Finally, we remove all control decisions which are not in $CP^\diamond$ to get an isolation supervisor $S_I^\diamond$ which is shown in Figure 9. From Figure 9, we can see that from each initial $Y$-state $y_0\in Y_0$, $S_I^\diamond$ will drive the system to states in $Y_m$ from which the type of faults can be determined.
	\begin{figure}[htb]
		\centering
		\includegraphics[scale = 0.6]{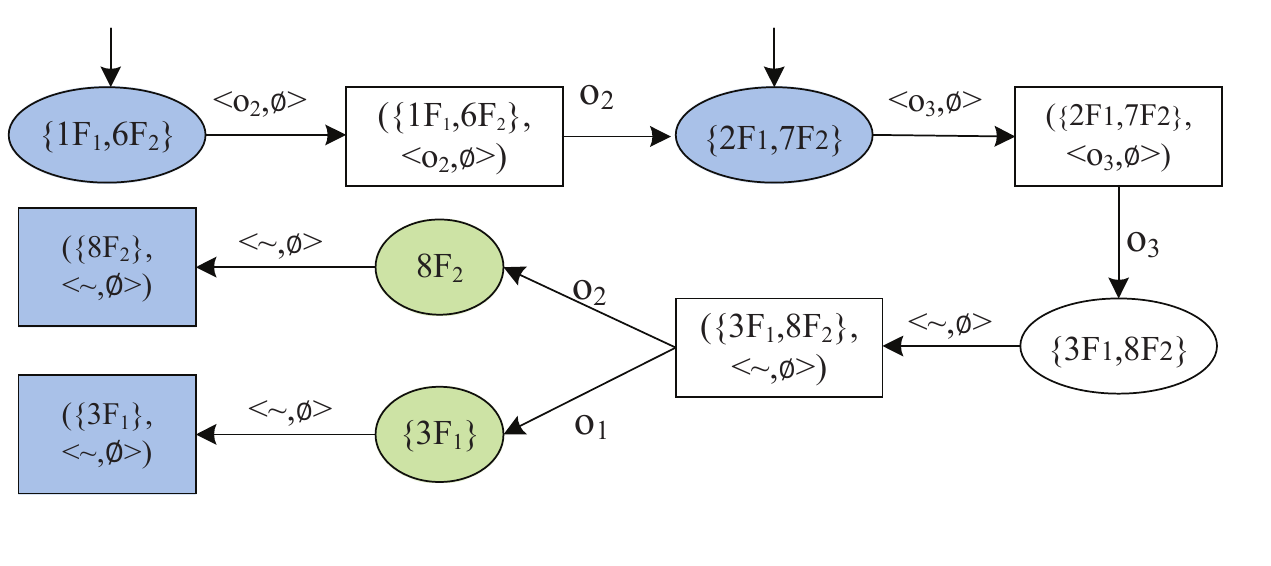}
		\caption{Valid isolation supervisor $S_I^\diamond$.}
		\label{Fig.9}
	\end{figure}		
\end{example}

\section{Applications on smart home}\label{Sec7}
In this section, we apply the results of active fault isolation to a smart home system to detect and isolate faults occurred in its lighting subsystem. The smart home system is used to control an office whose layout is shown in Figure 10. In the office, there are one left ceiling lamp $L$,  one right ceiling lamp $R$, and one floor lamp $F$. We can control their work status (on or off) via wireless network, respectively.  The light intensity is detected by an illuminance sensor.
\begin{figure}[htb]
	\centering
	\includegraphics[scale = 0.7]{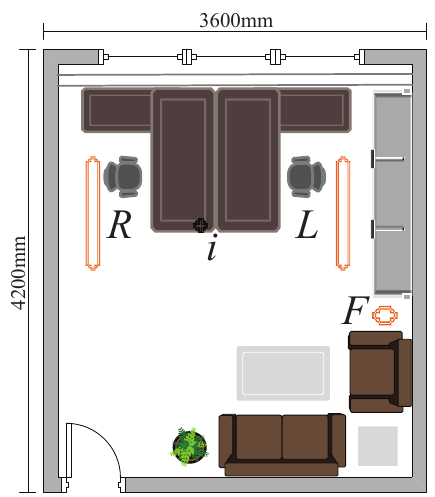}
	\caption{The layout of the office.}
	\label{Fig.10}
\end{figure}

When different lamps are turned on, the light intensity is also different. The data of light intensity is obtained through experiments for different cases. We abstract the continuous light intensity into 11 discrete sensor events $e_1,e_2,\cdots, e_{11}$. Each event represents that light intensity varies within specific upper bound and lower bound as shown in Figure 11.
\begin{figure}[htb]
	\centering
	\includegraphics[scale = 0.6]{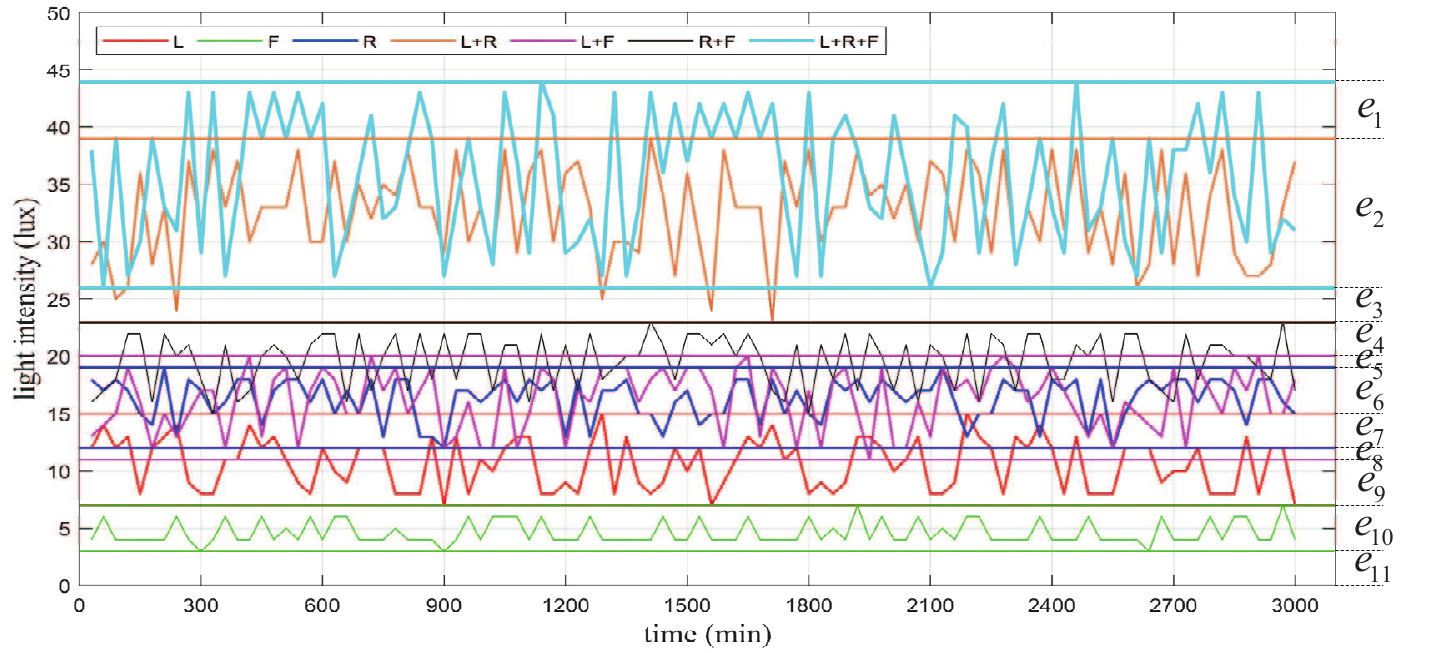}
	\caption{The light intensity of three lamps.}
	\label{Fig.11}
\end{figure}

The three lamps can be on or off. For each lamp, we consider the occurrence of breakdown failure. We then construct the automaton model for each lamp as shown in Figure 12.
\begin{figure}[htb]
	\centering
	\includegraphics[scale = 0.6]{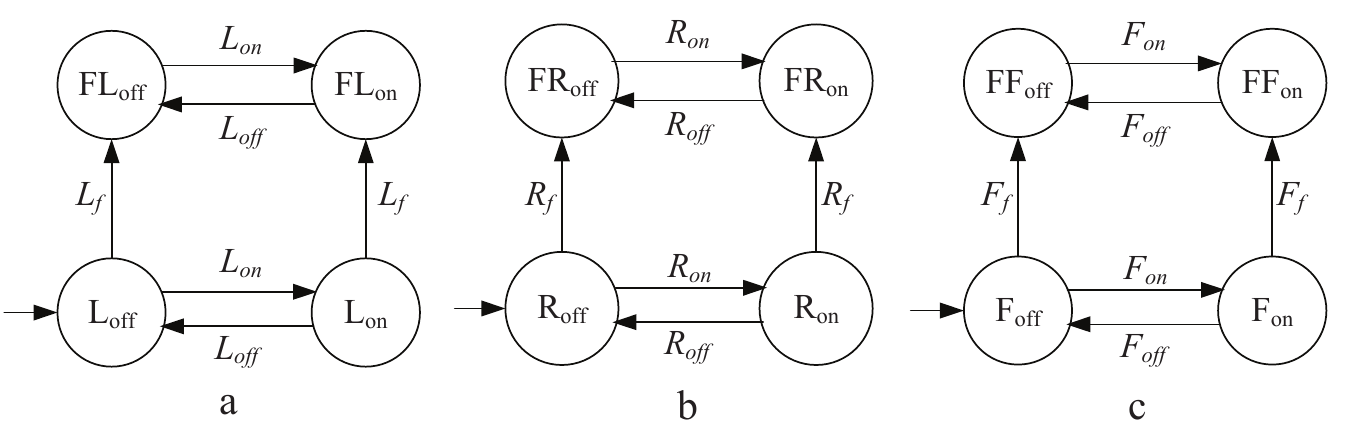}
	\caption{The model for three lamps: a. the left ceiling  lamp $G_a$, b. the right ceiling  lamp $G_b$ and c. the floor lamp $G_c$}
	\label{Fig.12}
\end{figure}
The events are defined as follows.
\begin{itemize}
	\item $L_{on}$: Turn on the left ceiling  lamp;
	\item $L_{off}$: Turn off the left ceiling  lamp;
	\item $R_{on}$: Turn on the right ceiling  lamp;
	\item $R_{off}$: Turn off the right ceiling  lamp;
	\item $F_{on}$: Turn on the floor lamp;
	\item $F_{off}$: Turn off the floor lamp;
	\item $L_f$: The left ceiling  lamp fails;
	\item $R_f$: The right ceiling  lamp fails;
	\item $F_f$: The floor lamp fails;
\end{itemize}

We construct the model of the lighting subsystem by parallel composition as $G_{abc}=G_{a}||G_{b}||G_{c}$ which has 32 states, 9 events and 116 transitions. For each state in $G_{abc}$, we obtain an actual range of light intensity through experiments. If the light intensity presented by a sensor event falls into the range, then the sensor event should occur at this state. In this way, we add sensor events for each state. Taking state 1 for example, the right ceiling lamp is turned on. From Figure 11, we know the light intensity varies from 12$lux$ to 19$lux$. The light intensity presented by sensor events $e_6$ and $e_7$ are both in the range. Hence from state 1, $e_6$ and $e_7$ can occur. We then obtain the complete model $G_{abc}^\circ$ for the lighting subsystem. Figure 13 shows part of $G_{abc}^\circ$. 

\begin{figure}[htb]
	\centering
	\includegraphics[scale = 0.75]{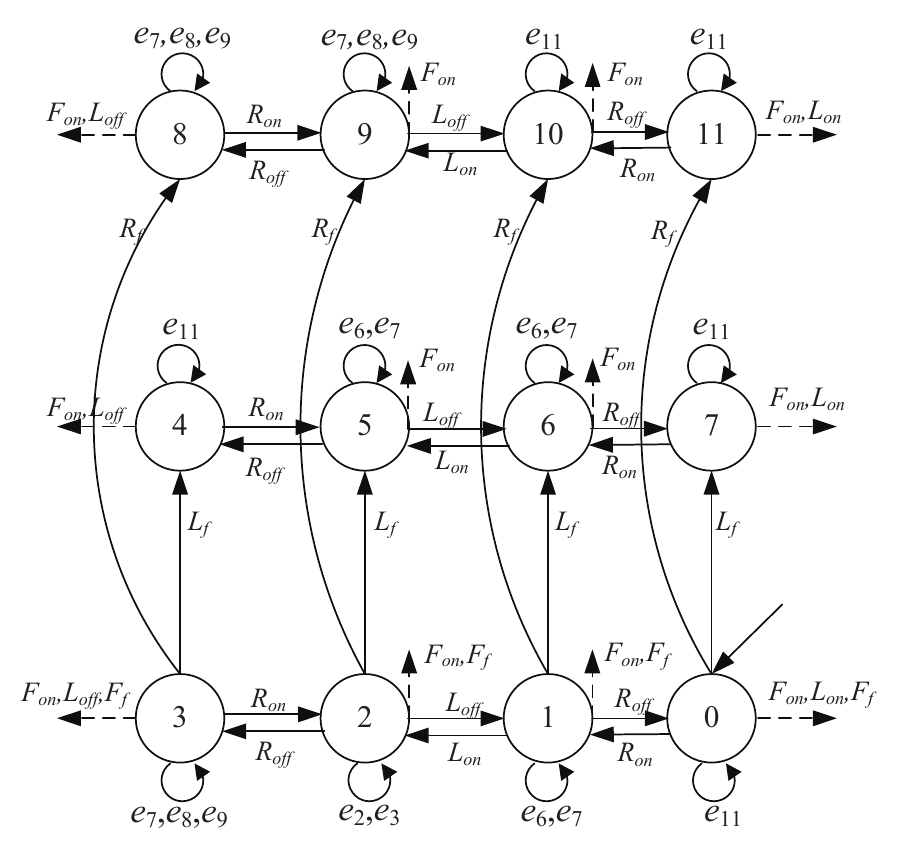}
	\caption{Part of $G_{abc}^\circ$}
	\label{Fig.13}
\end{figure}

For automaton $G_{abc}^\circ$, we have $\Sigma_{f_1}=\{L_F\}$, $\Sigma_{f_2}=\{R_f\}$, $\Sigma_{f_3}=\{F_f\}$, $\Sigma_{uo}=\{L_f,R_f,F_f\}$ and
\begin{align*}
\Sigma_{o}=\{L_{on},L_{off},R_{on},R_{off},F_{on},F_{off},e_1,e_2,\cdots,e_{11}\}.
\end{align*}
Let us construct its diagnoser $G_d=(X,\Sigma_o,\xi,x_0)$ which has 51 states, 17 events and 317 state transitions. Figure 14 shows part of $G_d$.
\begin{figure}[htb]
	\centering
	\includegraphics[scale = 0.6]{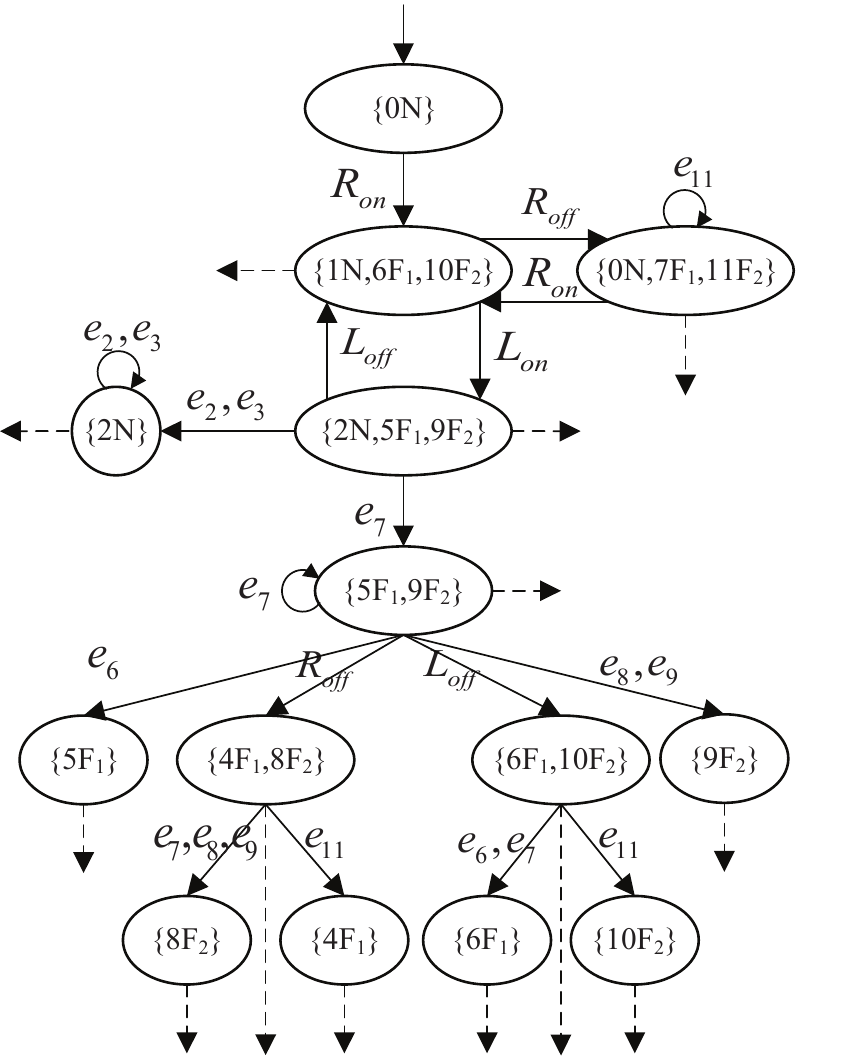}
	\caption{Part of $G_d$}
	\label{Fig.14}
\end{figure}

From Figure 14 , we can find that the system is not isolatable. For example, when the diagnoser reaches $\{5F_1,9F_2\}\in X_{F,0}$, we can not determine which type of fault has occurred with more observations. Let us control the system to be isolatable. In the lighting system, the forcible event set is equal to the controllable event set as $\Sigma_{en}=\Sigma_c=\{L_{on},L_{off},R_{on},R_{off},F_{on},F_{off}\}$. For $\{5F_1,9F_2\}\in X_{F,0}$, we set $Y_0=\{\{5F_1,9F_2\}\}$ and then construct the bipartite transition system $BTS$ for the lighting system to obtain all feasible isolation supervisors. With Algorithm 1 and Algorithm 2, we can verify $Y_0\subseteq Y_g$. Hence we can control the system to determine the type of faults. By Algorithm 2, we can derive one valid isolation supervisor ${S_I^{\diamond}}$. Figure 15 shows part of ${S_I^{\diamond}}$.
\begin{figure}[htb]
	\centering
	\includegraphics[scale = 0.6]{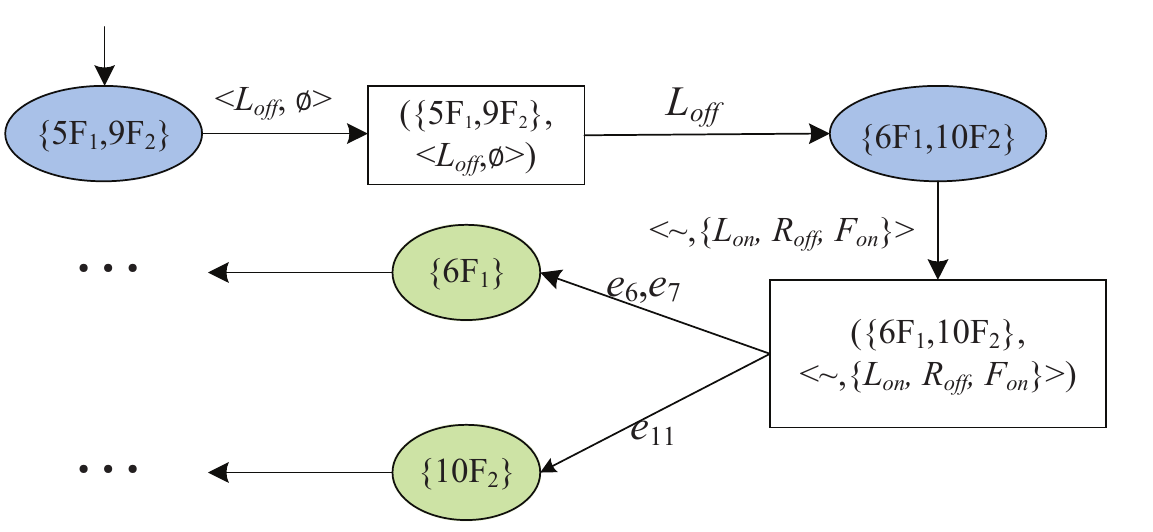}
	\caption{Part of ${S_I^{\diamond}}$}
	\label{Fig.15}
\end{figure}

The valid isolation supervisor ${S_I^{\diamond}}$ shown in Figure 15 works as follows. For $Y$-state $\{5F_1,9F_2\}$, $5F_1$ means two ceiling lamps are turned on and the left ceiling lamp has failed and $9F_2$ means two ceiling lamps are turned on and the right ceiling lamp has failed. In order to determine the specific fault type, we close the left ceiling lamp. We then keep only the right ceiling lamp open by disabling events $L_{on}$, $L_{off}$ and $F_{on}$ (prohibiting turning on the left ceiling lamp, turning off the right ceiling lamp and turning on the floor lamp). Finally, by observing the sensor events, if the light intensity approaches to zero (observing the occurrence of event $e_{11}$), we know the right ceiling lamp has failed. Otherwise, we will see $e_6$ or $e_7$ which means the right ceiling lamp works normally and hence the left ceiling lamp has failed.

\section{Conclusions}\label{Sec8}
In this paper, we investigate the active fault isolation problem for discrete event systems. By combining two different control mechanisms of disabling controllable events and enforcing forcible events, we successfully synthesize a new supervisor to isolate faults. More specifically, we construct a bipartite transition system that embeds all feasible isolation supervisors and derive a necessary and sufficient condition for the existence of solutions to the active fault isolation problem. We also develop two algorithms to verify the condition and construct valid isolation supervisors.

In the future, we plan to extend our results on active fault isolation to distributed discrete event systems and use the method of combining two different control mechanisms to synthesize more powerful supervisors for other supervisory control problems.

\bibliographystyle{IEEEtran}
\bibliography{reference}

\begin{thebibliography}{10}
\providecommand{\url}[1]{#1}
\csname url@samestyle\endcsname
\providecommand{\newblock}{\relax}
\providecommand{\bibinfo}[2]{#2}
\providecommand{\BIBentrySTDinterwordspacing}{\spaceskip=0pt\relax}
\providecommand{\BIBentryALTinterwordstretchfactor}{4}
\providecommand{\BIBentryALTinterwordspacing}{\spaceskip=\fontdimen2\font plus
\BIBentryALTinterwordstretchfactor\fontdimen3\font minus
  \fontdimen4\font\relax}
\providecommand{\BIBforeignlanguage}[2]{{%
\expandafter\ifx\csname l@#1\endcsname\relax
\typeout{** WARNING: IEEEtran.bst: No hyphenation pattern has been}%
\typeout{** loaded for the language `#1'. Using the pattern for}%
\typeout{** the default language instead.}%
\else
\language=\csname l@#1\endcsname
\fi
#2}}
\providecommand{\BIBdecl}{\relax}
\BIBdecl

\bibitem{2015ASurvey}
Z.~Gao, C.~Cecati, and S.~X. Ding, ``A survey of fault diagnosis and
  fault-tolerant techniques—part i: Fault diagnosis with model-based and
  signal-based approaches,'' \emph{IEEE Transactions on Industrial
  Electronics}, vol.~62, no.~6, pp. 3757--3767, 2015.

\bibitem{1994Real}
A.~D. Pouliezos and G.~S. Stavrakakis, \emph{Real Time Fault Monitoring of
  Industrial Processes}.\hskip 1em plus 0.5em minus 0.4em\relax Kluwer Academic
  Publishers, 1994.

\bibitem{2010Asurvey}
I.~Hwang, S.~Kim, Y.~Kim, and C.~E. Seah, ``A survey of fault detection,
  isolation, and reconfiguration methods,'' \emph{IEEE Transactions on Control
  Systems Technology}, vol.~18, no.~3, pp. 636--653, 2010.

\bibitem{Cassandras2010Introduction}
C.~G. Cassandras and S.~Lafortune, \emph{Introduction to Discrete Event
  Systems}.\hskip 1em plus 0.5em minus 0.4em\relax Springer US, 2010.

\bibitem{2013Overview}
J.~Zaytoon and S.~Lafortune, ``Overview of fault diagnosis methods for discrete
  event systems,'' \emph{Annual Reviews in Control}, vol.~37, no.~2, p.
  308–320, 2013.

\bibitem{2018On}
S.~Lafortune, F.~Lin, and C.~N. Hadjicostis, ``On the history of diagnosability
  and opacity in discrete event systems,'' \emph{Annual Reviews in Control},
  vol.~45, pp. 257--266, 2018.

\bibitem{lin1994diagnosability}
F.~Lin, ``Diagnosability of discrete event systems and its applications,''
  \emph{Discrete Event Dynamic Systems}, vol.~4, no.~2, pp. 197--212, 1994.

\bibitem{SaSe:95}
M.~Sampath, R.~Sengupta, S.~Lafortune, K.~Sinnamohideen, and D.~Teneketzis,
  ``Diagnosability of discrete-event systems,'' \emph{IEEE Transactions on
  Automatic Control}, vol.~40, no.~9, pp. 1555--1575, 1995.

\bibitem{CaBaMo:12}
L.~K. Carvalho, J.~C. Basilio, and M.~V. Moreira, ``Robust diagnosis of
  discrete event systems against intermittent loss of observations,''
  \emph{Automatica}, vol.~48, no.~9, pp. 2068--2078, 2012.

\bibitem{CaMoBaLa:13}
L.~K. Carvalho, M.~V. Moreira, J.~C. Basilio, and S.~Lafortune, ``Robust
  diagnosis of discrete-event systems against permanent loss of observations,''
  \emph{Automatica}, vol.~49, no.~1, pp. 223--231, 2013.

\bibitem{2021Cao}
L.~Cao, S.~Shu, F.~Lin, Q.~Chen, and C.~Liu, ``Weak diagnosability of discrete
  event systems,'' \emph{IEEE Transactions on Control of Network Systems},
  2021.

\bibitem{JiHuChKu:01}
S.~Jiang, Z.~Huang, V.~Chandra, and R.~Kumar, ``A polynomial algorithm for
  testing diagnosability of discrete-event systems,'' \emph{IEEE Transactions
  on Automatic Control}, vol.~46, no.~8, pp. 1318--1321, 2001.

\bibitem{YoLa:02}
T.~S. Yoo and S.~Lafortune, ``Polynomial-time verification of diagnosability of
  partially observed discrete-event systems,'' \emph{IEEE Transactions on
  Automatic Control}, vol.~47, no.~9, pp. 1491--1495, 2002.

\bibitem{sampath1996failure}
M.~Sampath, R.~Sengupta, S.~Lafortune, K.~Sinnamohideen, and D.~C. Teneketzis,
  ``Failure diagnosis using discrete-event models,'' \emph{IEEE transactions on
  control systems technology}, vol.~4, no.~2, pp. 105--124, 1996.

\bibitem{ThTe:05}
D.~{Thorsley} and D.~{Teneketzis}, ``Diagnosability of stochastic
  discrete-event systems,'' \emph{IEEE Transactions on Automatic Control},
  vol.~50, no.~4, pp. 476--492, 2005.

\bibitem{2013Diagnosis}
M.~P. Cabasino, A.~Giua, and C.~Seatzu, ``Diagnosis using labeled petri nets
  with silent or undistinguishable fault events,'' \emph{IEEE Transactions on
  Systems, Man, and Cybernetics: Systems}, vol.~43, no.~2, p. 345–355, 2013.

\bibitem{2005Decentralized}
W.~Qiu and R.~Kumar, ``Decentralized failure diagnosis of discrete event
  systems,'' \emph{IEEE Transactions on Systems, Man, and Cybernetics: Systems
  and Humans}, vol.~36, no.~2, pp. 384--395, 2005.

\bibitem{2013Reliable}
S.~Nakata and S.~Takai, ``Reliable decentralized failure diagnosis of discrete
  event systems,'' \emph{Sice Journal of Control Measurement \& System
  Integration}, vol.~6, no.~5, pp. 353--359, 2013.

\bibitem{2002Fault}
S.~H. Zad, R.~H. Kwong, and W.~M. Wonham, ``Fault diagnosis in timed
  discrete-event systems,'' in \emph{IEEE Conference on Decision \& Control},
  2002.

\bibitem{2010The}
F.~Cassez, ``The complexity of codiagnosability for discrete event and timed
  systems,'' \emph{Springer Berlin Heidelberg}, 2010.

\bibitem{1998Active}
M.~Sampath, S.~Lafortune, and D.~Teneketzis, ``Active diagnosis of
  discrete-event systems,'' \emph{IEEE Transactions on Automatic Control},
  vol.~43, no.~7, pp. 908--929, 1998.

\bibitem{2014Active}
N.~Bertrand, R.~Fabre, S.~Haar, S.~Haddad, and H.~Loc, ``Active diagnosis for
  probabilistic systems,'' \emph{International Conference on Foundations of
  Software Science and Computation Structures}, 2014.

\bibitem{Stefan2017Optimal}
S.~Haar, S.~Haddad, T.~Melliti, and S.~Schwoon, ``Optimal constructions for
  active diagnosis,'' \emph{Journal of Computer \& System Sciences}, 2017.

\bibitem{2016A}
X.~Yin and S.~Lafortune, ``A uniform approach for synthesizing
  property-enforcing supervisors for partially-observed discrete-event
  systems,'' \emph{IEEE Transactions on Automatic Control}, vol.~61, no.~8, pp.
  2140--2154, 2016.

\bibitem{2020Design}
Y.~{Hu}, Z.~{Ma}, and Z.~{Li}, ``Design of supervisors for active diagnosis in
  discrete event systems,'' \emph{IEEE Transactions on Automatic Control},
  vol.~65, no.~12, pp. 5159--5172, 2020.

\bibitem{2005Control}
T.~Ushio and S.~Takai, ``Control-invariance of hybrid systems with forcible
  events,'' \emph{Automatica}, vol.~41, no.~4, pp. 669--675, 2005.

\bibitem{2020On}
F.~{Lin}, L.~{Wang}, W.~{Chen}, and M.~P. {Polis}, ``On controllability of
  hybrid systems,'' \emph{IEEE Transactions on Automatic Control}, vol.~66,
  no.~7, pp. 3243--3250, 2021.

\bibitem{2014ActiveD}
Z.~Chen, F.~Lin, C.~Wang, L.~Wang, and M.~Xu, ``Active diagnosability of
  discrete event systems and its application to battery fault diagnosis,''
  \emph{IEEE Transactions on Control Systems Technology}, vol.~22, no.~5, pp.
  1892--1898, 2014.

\bibitem{2017N}
F.~Lin, L.~Y. Wang, W.~Chen, L.~Han, and B.~Shen, ``N -diagnosability for
  active on-line diagnosis in discrete event systems,'' \emph{Automatica},
  vol.~83, pp. 220--225, 2017.

\bibitem{YiLa16}
X.~Yin and S.~Lafortune, ``Synthesis of maximally permissive supervisors for
  partially observed discrete event systems,'' \emph{IEEE Transactions on
  Automatic Control}, 2016.

\end{thebibliography}
\begin{IEEEbiography}[{\includegraphics[width=1in,height=1.25in,clip,keepaspectratio]{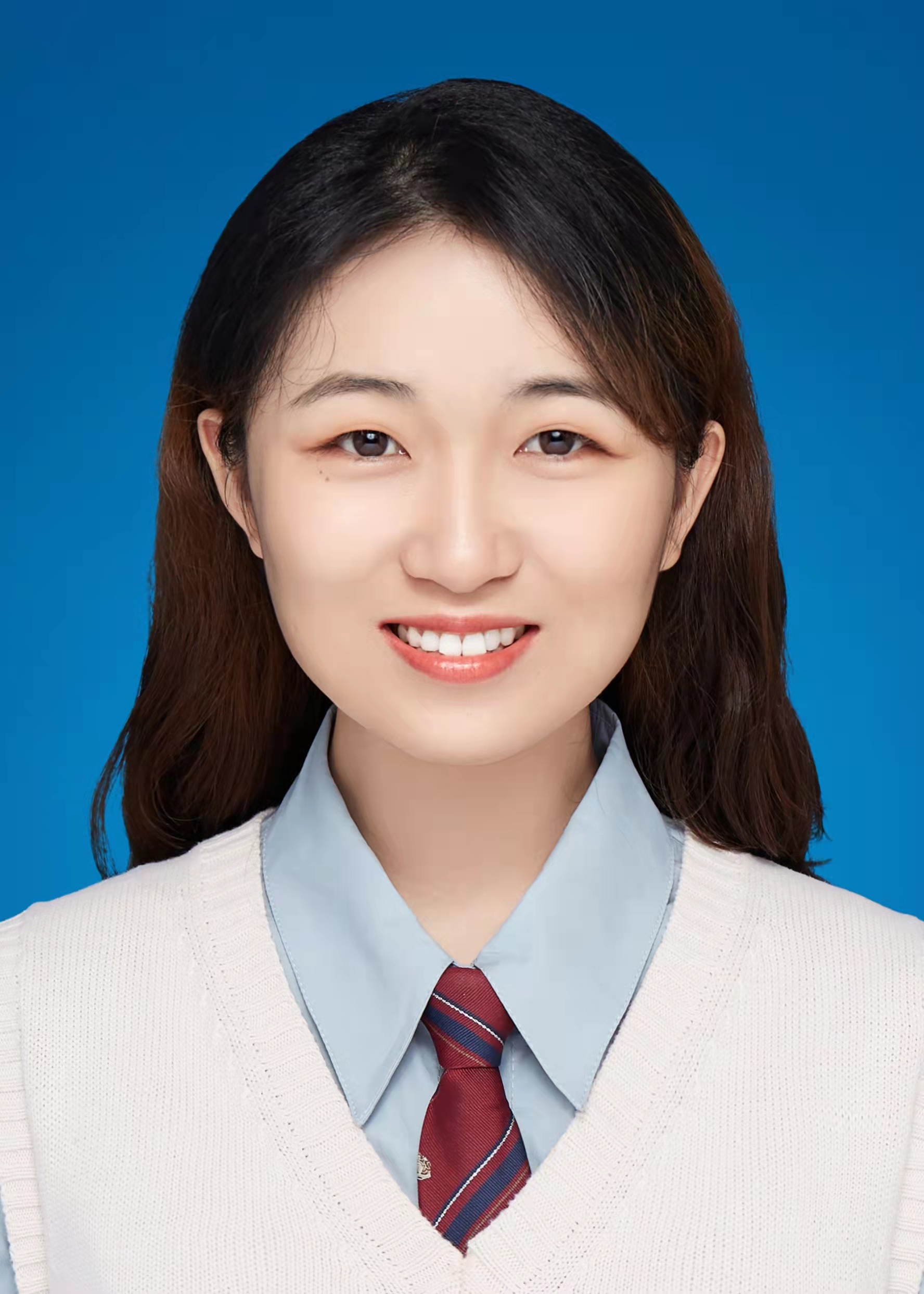}}]{Lin Cao}
	was born in Anhui, China, in 1996. She received her B.Eng.
	degree in process equipment and control engineering from Dalian University of Technology,
	Liaoning, China, in 2018. Since September,
	2018, she has been with the School of Electronics and
	Information Engineering, Tongji University, Shanghai, China,
	where she is currently a Ph.D. candidate.
	Her main research interests include control of discrete event systems and cyber-physical systems.
\end{IEEEbiography}

\begin{IEEEbiography}[{\includegraphics[width=1in,height=1.25in,clip,keepaspectratio]{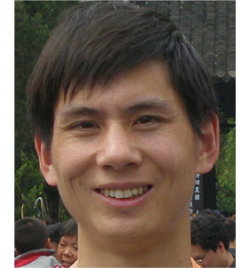}}]{Shaolong Shu}
	was born in Hubei, China, in 1980. He received his B.Eng. degree in automatic control, and his Ph.D. degree in control theory and control engineering from Tongji University, Shanghai, China, in 2003 and 2008, respectively. Since July, 2008, he has been with the School of Electronics and Information Engineering, Tongji University, Shanghai, China, where he is currently a professor. From August, 2007 to February, 2008 and from April, 2014 to April, 2015, he was a visiting scholar in Wayne State University, Detroit, MI, USA. His main research interests include state estimation, fault-tolerant control and networked control of discrete event systems.
\end{IEEEbiography}

\begin{IEEEbiography}[{\includegraphics[width=1in,height=1.25in,clip,keepaspectratio]{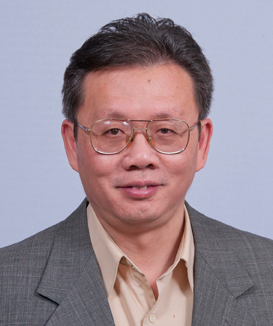}}]{Feng Lin}
	(S85-M88-SM07-F09) received his B.Eng. degree in electrical engineering from Shanghai Jiao Tong University,	Shanghai, China, in 1982, and the M.A.Sc. and Ph.D. degrees in electrical engineering from the University of Toronto, Toronto,
	ON, Canada, in 1984 and 1988, respectively. He was a Post-Doctoral Fellow with Harvard University, Cambridge, MA, USA, from 1987 to 1988. Since 1988, he has been with the Department of Electrical and Computer Engineering, Wayne State University, Detroit, MI, USA, where he is currently a Professor. His current research interests include discrete event systems, hybrid systems, robust control, and their applications in alternative energy, biomedical systems, and automotive control. He authored a book entitled Robust Control Design: An Optimal Control Approach and coauthored a paper that received a George Axelby outstanding paper award from the IEEE Control Systems Society. He was an associate editor of
	IEEE Transactions on Automatic Control.
\end{IEEEbiography}

\end{document}